\documentclass[12pt,preprint]{aastex}

\usepackage{graphicx}

\slugcomment{to appear in ARAA 2007}

\shorttitle{Irregular Satellites}
\shortauthors{Jewitt and Haghighipour}

\begin{document}

\title{Irregular Satellites of the Planets: \\
Products of Capture in the Early Solar System}

\author{David Jewitt}

\and

\author{Nader Haghighipour}
\affil{Institute for Astronomy, University of Hawaii, \\
2680 Woodlawn Drive, Honolulu, HI 96822}

\email{jewitt@hawaii.edu, nader@ifa.hawaii.edu}

\begin{abstract}
All four giant planets in the Solar system possess irregular satellites, 
characterized by large, highly eccentric and/or inclined orbits that are 
distinct from the nearly circular, uninclined orbits of the regular 
satellites. 
This difference can be traced directly to different modes of formation. 
Whereas the regular satellites grew by accretion within circumplanetary 
disks the irregular satellites were captured from initially heliocentric 
orbits at an early epoch. Recently, powerful survey observations have 
greatly increased the number of known irregular satellites, permitting 
a fresh look at the group properties of these objects and motivating a 
re-examination of the mechanisms of capture. None of the suggested mechanisms, 
including gas-drag, pull-down, and three-body capture, convincingly fit 
the group characteristics of the irregular satellites. The sources of the 
satellites also remain unidentified. 
\end{abstract}


\section{Definition}

Planetary satellites are naturally divided on the basis of their 
orbits into two distinct classes. Qualitatively, the so-called 
regular satellites are confined to the central portions 
(typically less than a few percent) of their planets' Hill spheres. 
The Hill sphere is the domain over which a planet exerts 
gravitational control in competition with the Sun. It corresponds 
roughly to the size of the more familiar Roche lobe surrounding 
each planet, and has a radius

\begin{equation}
r_H \sim  a_p \left(\frac{\mu}{3}\right)^{1/3}.
\label{hill}
\end{equation}

\noindent 
Here, $a_p$ is the orbital semimajor axis of the planet 
and $\mu$ = ${m_p}/{M_{\odot}}$, where $m_p$ and $M_{\odot}$ 
are the masses of the planet and Sun, respectively.  
Values of $r_H$ for the giant planets are from $\sim$0.35 AU 
to 0.77 AU, increasing with distance from the Sun (Table 1). Most  
regular satellites follow orbits of low eccentricity 
$(\leq 0.01)$ and small inclination (a few degrees).  
In contrast, the ``irregular satellites'' have orbit sizes 
that extend up to $\sim$0.5 r$_H$ and their 
eccentricities and inclinations are commonly large 
($\sim$0.1 to $\sim$0.7 and up to 180$^{\circ}$, respectively). 

Other definitions have been invoked to distinguish irregular 
satellites from
regular satellites.  For example, Burns (1986) defined satellites 
as irregular when their orbital planes precess primarily under the
influence of torques from the Sun (rather than from the oblate 
planets).    This definition leads to a critical semimajor axis 
for orbits
about each planet, given by 

\begin{equation}
a_c \sim (2 \mu J_2 R_e^2 a_p^3)^{1/5}
\end{equation}

\noindent 
in which $J_2$ is the second spherical harmonic
(describing the planet's oblateness), $R_e$ is the planetary 
equatorial radius and the other variables are as defined above. 
Satellites with $a > a_c$ are classified as irregular.  
Practically, the distinction between regular and irregular 
satellites is relatively sharp, and the different definitions 
give the same result.  The
main exception is Neptune's large satellite Triton, which is 
excluded by the precession criterion because its orbit is small 
and relatively immune to Solar perturbations.  As we
discuss later, there are good reasons to believe that Triton 
should be grouped with the irregular satellites (not least 
because its orbit is retrograde) but its large size and small 
orbit separate it from the other 
irregulars in important ways.   By either definition, 
about 100 irregular satellites are known.

This review is motivated by recent developments in the study 
of irregular planetary satellites.  Use of large-format 
charge-coupled device (CCD) detectors has powered an 
unprecedented wave of irregular satellite discoveries 
and theoretical interest
in the origin and significance of these bodies has 
likewise intensified.  The irregular satellites were reviewed by
Cruikshank, Degewij \& Zellner  (1982), when 
only $\sim$10 such bodies 
were known.  Their connections to the Trojans and to 
temporary satellites were discussed in 
Jewitt et al. (2004) and we draw attention to a 
popular-level description (Jewitt et al. 2006).

\subsection{Why Do They Matter?}

Regular satellites were formed in the equatorial accretion
disks of their host planets (Lunine \& Stevenson 1982; 
Canup \& Ward 2002, 2006; Mosqueira \& Estrada 2003) 
but this is not a viable explanation for the irregular 
satellites. In particular, many irregular satellites follow 
retrograde orbits (inclinations $>$ 90°) that are incompatible 
with formation in prograde rotating accretion disks. 
The most plausible explanation is that the irregular 
satellites were captured by the planets from orbits that 
were initially heliocentric. This difference in the modes 
of formation is what conveys fundamental importance to 
the study of the irregular satellites.

Temporary captures of passing bodies by planets are common 
(Carusi \& Valsecchi 1979). A famous example is the temporary 
capture of comet D/Shoemaker-Levy 9, which ended dramatically 
with the impact of the comet into Jupiter (Weaver et al. 1995). 
Planetary impacts like that of D/Shoemaker-Levy 9 occur with 
a $\sim$1000 year timescale, but a more usual fate is for 
temporary captures to last for a few tens of years and to 
be terminated by the escape of the trapped body back into 
heliocentric orbit (Benner \& McKinnon 1995, Kary \& Dones 1996). 
Permanent capture of a body from heliocentric orbit into a 
bound, planetocentric orbit requires the action of some 
nonconservative process, for example, frictional dissipation 
or energy loss through collisions. The modern-day Solar 
system offers no such process. Therefore, the capture of 
the irregular satellites is presumed to have occurred at 
early times, when the gross properties of the Solar system 
may have been different from those that now prevail. 
Capture could have occurred in association with planet 
formation in the presence of residual gas, or at a later 
stage corresponding to the final clearing of the outer 
Solar system. In any event, the scientific importance of 
the irregular satellites lies in their capacity to tell 
us about capture processes in the early Solar system: 
the irregular satellites may provide a window onto 
otherwise unobserved times.

\section{Observational Background} 

Most planetary satellites were discovered using one of 
three different forms of detector technology. The brightest 
and first-discovered examples were found telescopically 
by intrepid visual observers of old, starting with Galileo's 
discovery of the four giant satellites of Jupiter in 1610. 
Almost all of the early discoveries were of regular satellites. 
The second wave of discovery relied on photographic plates 
to provide wide coverage of the sky at higher sensitivity 
than possible by eye. For a while, it was common practice 
for observatory directors to prove the worth of major new 
telescopes by using them to discover a planetary satellite 
or two (Kuiper 1961). The improved sensitivity of the 
photographic surveys over the human eye uncovered a growing 
number of irregular satellites. By the end of the twentieth 
century about 10 such objects were known (Figure 1). 
The third wave of satellite discovery, and the one that 
continues now, employs large-format CCD imagers on large 
telescopes to survey the planetary Hill spheres to even 
greater depths. These modern CCD surveys have, in the past 
half decade, increased the number of known irregular 
satellites by an order of magnitude to about 100 (Figure 1), 
showing that these objects are probably numerically dominant 
over (but systematically smaller than) the regular satellites. 
The improved satellite samples are beginning to reveal the 
global properties of the irregular satellite systems of 
different planets and have provided motivation for a number 
of exciting theoretical investigations into their dynamics 
and possible origins. The third wave of discovery is also 
the driver for this review.

The inverse square law connects the heliocentric and geocentric 
distances, $R$ (AU) and $\Delta$ (AU), of the satellite to its 
apparent magnitude, $m_R$:

\begin{equation}
p_R r^2 = 2.25 \times 10^{22} R^2 \Delta^2 10^{0.4(m_{\odot} - m_R)}.
\label{inverse}
\end{equation}

\noindent 
where $r$ (km), is the radius of the satellite, $p_R$ is its
geometric albedo, and $m_\odot$ is the apparent magnitude
of the Sun. At opposition, $\Delta$ = $R$ - 1.  
With $R \gg$ 1 and substituting $p_R$ = 0.04, this relation gives 

\begin{equation}
r~[km] \sim \left[\frac{R}{5}\right]^2 10^{0.2(24-m_R)}.  
\end{equation}

\noindent 
For example, Equation (3), and Figure \ref{sizes}, show that 
satellite surveys made to magnitude $m_R$ = 24 reach limiting 
radii $r \sim$ 1, 4, 16 and 36 km at Jupiter, Saturn, Uranus 
and Neptune, respectively. Relative to Jupiter, satellites of a
given size and albedo will be fainter at Saturn, Uranus and 
Neptune by 2.6, 5.9 and 7.6 magnitudes, respectively
(Table 1).  For this reason we know of a large number of 
(mostly small) irregular satellites at Jupiter but only 
smaller numbers of larger objects at the other giant planets
(Table 2).

\section{Properties of the Irregular Satellite Populations}

Most twentieth century surveys in which irregular satellites 
were discovered were conducted using photographic plates and, 
by modern standards, they are not well characterized. Indeed, 
the circumstances of a majority of these discoveries are not 
even published, and the most scientifically useful description 
of this early work is the summary by Kuiper (1961). The use 
of CCDs in the surveys of the past decade has made it easier 
to assess the limiting magnitude and effective area of each 
survey. These quantities are listed in Table 3 for the major, 
published irregular satellite-producing surveys.

The orbital characteristics of the known irregular satellites 
are summarized graphically in Figures 3 and 4. Figure 3 shows 
the orbital semimajor axis (normalized to the Hill sphere radius) 
plotted against the orbital inclination. Figure 4 is the 
corresponding plot against orbital eccentricity. 

The data from Figures 3 and 4 are shown in a different way in 
Figure 5. In this Figure each satellite is represented by a 
point whose distance 
from the origin gives the semimajor axis in units of the Hill sphere. 
The angle from the x-axis to each point indicates the inclination and 
the eccentricity is represented by the length of the bar on each point.
 From Figures 3, 4, and 5 the following general characteristics of the 
satellite orbits may be discerned:

\begin{itemize}

\item Retrograde satellites ($i >$ 90$^{\circ}$) outnumber 
prograde satellites at each planet
(Figures \ref{avsi} and \ref{aeiplot}). Overall, the ratio 
retrograde/prograde is 88/19 $\sim$4.5 (Table \ref{surveys}).
No known observational 
bias can produce such an asymmetry. Instead, it must result from 
either an asymmetry in the capture efficiency or greater 
dynamical/collisional stability of the retrograde satellites, 
or some combination of these effects. As we discuss below, models 
of the capture process tend to be symmetric with respect to 
inclination, so the asymmetry is more likely to reflect greater 
long-term stability of the retrograde satellites.

\item The retrograde satellites  ($x <$ 0 in Figure \ref{aeiplot}) 
have semimajor axes and eccentricities that are systematically 
larger than those of prograde satellites. This probably reflects 
greater stability of the retrograde satellites, which can orbit 
at greater distances without being lost from their planets.

\item  The semimajor axes are spread over a wide range with 
a maximum near $a/ r_H \sim$ 0.5 (Figure \ref{avse}).
It is true that most published
 surveys are biased toward the inner portions of the Hill spheres 
leading to the suspicion that more distant satellites might have
 been missed. This is especially true of the Jupiter and Saturn 
systems, where the large angular size subtended by
$r_H$ (Table \ref{hilltable}) 
is a major challenge to the surveys. However, with the large
 eccentricities characteristic of the irregular satellites, 
even objects with $a/r_H >$ 0.5 would have periapses 
in the surveyed regions, 
and so would have a finite probability of being detected.
 Only distant, low-eccentricity satellites might have been 
missed by some surveys. It seems safe to conclude that the
 outer half of the Hill sphere of each planet is greatly depleted 
in satellites relative to the inner half.

\item The median values of the normalized semimajor axes are 
$a/ r_H$ = 0.44, 0.29, 0.17, and 0.19, for Jupiter, Saturn, Uranus,
and Neptune, respectively (Figures \ref{avsi} and \ref{avse}).
 This trend toward 
smaller satellite systems around the more distant planets is not 
likely an artifact of survey bias (which, if present, would tend 
to produce an opposite trend). Neither is it an expected 
consequence of long-term dynamical instability.
Nesvorn\'y et al. (2003) noted that the satellites of the outer
planets would be destroyed by mutual collisions in the lifetime
of the Solar system if displaced to orbits around Jupiter.
On this basis, they assert that the $a/r_H$ versus $a_p$   
trend could be a result of past collisional depletion.

\item No irregular satellites have been found with inclinations 
in the range 60 $\le i \le$ 130$^{\circ}$. 
The polar regions have been surveyed, 
and this absence is not an artifact of observational bias. Instead, 
the lack of highly inclined orbits most likely reflects an 
instability induced by the Kozai resonance, discussed in Section 5. 

\item The Jovian irregulars are clustered in $a$/$r_H$  
versus $i$ space. 
Major clusters (or ``families'') are labeled in Figure \ref{avsi}  
with the
names of the largest members (from Sheppard \& Jewitt 2003, 
also Nesvorny et al. 2003). Relative velocities among family members
 are comparable to the escape velocity from the largest member 
(e.g., 100 m s$^{-1}$ for a 100 km-scale body). The Saturnian 
irregulars may also be clustered in inclination alone 
(e.g., see the set of four prograde satellites with 
$i \sim$ 45$^{\circ}$ 
spread over 0.17 $\le a/r_H \le$ 0.28 in Figure \ref{avsi}). 
However, the Saturn families 
are not tight in $a$/$r_H$ versus $i$ like those at Jupiter. 
The Uranian 
and Neptunian satellites are too few in number for any meaningful 
statement about clustering.

\item Although the satellites are distributed non-randomly in 
the $a$/$r_H$ versus $e$ plane (Figure \ref{avse}), evidence for 
tight clustering is much less evident than in $a$/$r_H$ versus $i$.  
For example, the tight Himalia group in Figure \ref{avsi} is only a 
loose assemblage in Figure \ref{avse}.   

\end{itemize}

\subsection{Physical Properties}
\subsubsection{Colors}

Optical color measurements (Smith et al. 1981; Tholen \& Zellner 1984; 
Luu 1991; Rettig et al. 2001; Grav et al. 2003, 2004a,b; 
Grav \& Bauer 2007) show that the irregular satellite surfaces 
vary from neutral (Sun-colored) to moderately red. The most
 reliable color measurements, those having 1$\sigma$  uncertainties 
smaller than 10\%, are plotted in Figure \ref{color_color}, 
where they are 
compared with the colors of Kuiper Belt objects (KBOs). One 
conclusion we can drawn from Figure \ref{color_color},
 is that the colors of 
the irregular satellite populations of the different planets 
are indistinguishable. This is consistent with (but does not prove) 
a common origin for the irregular satellites, as would be 
expected if they were captured from a common source.

Another conclusion is that the satellite colors are, on average, 
systematically bluer than the colors of the KBOs. Specifically, 
Figure  \ref{color_color}  shows that this is because the 
satellites are 
(with the possible exception of Jupiter's satellite XXIII Kalyke) 
lacking in the ``ultrared matter'' (Jewitt 2002) that characterizes 
many of the KBOs. By definition, ultrared matter has a spectral 
reflectivity that increases with wavelength by more than 
25\% per 1000\AA. It is probably an indicator of the presence 
of surface organics, since most cosmochemically plausible inorganic 
materials are less red. The ultrared matter is not found in the 
small-body populations of the inner solar system, perhaps as a 
result of its ejection or burial by sublimation-driven outgassing
(Jewitt 2002). Likewise, organics on irregular satellites of 
Jupiter (which, at  $\sim$5 AU, lies at the outer edge of
 the water ice 
sublimation zone) might have been ejected or buried by past 
activity. However, the same explanation is less viable on the 
irregular satellites of the more distant planets, since these 
are too cold for sublimation to occur. If the color systematics 
in Figure \ref{color_color}   survive the addition of new data, 
then the absence 
of ultrared objects will be an important constraint on the possible 
source regions from which irregular satellites are captured.

The colors of satellites within dynamically defined families are,
in general, more similar to each other than they are to the members 
of other families (Grav et al. 2003). This is consistent with the 
contention that the satellites within families are fragments of a 
single, homogeneous parent, although space weathering may act to 
produce spectral uniformity as observed.
Beyond broadband color measurements, few spectra of the irregular 
satellites exist. The bright irregulars J VI Himalia and S IX 
Phoebe have been studied in detail. They are, respectively, 
spectrally featureless and dominated by the bands of water ice
(see Section 6).

\subsubsection{Size distributions}

The brightness of a body viewed in scattered light is related to 
the product of the cross-sectional area with the geometric albedo 
measured at the wavelength of observation (Equation  \ref{inverse}).
For most 
irregular satellites we lack independent measurements of the albedo, 
and so the effective areas, and hence sizes, of the satellites can 
be determined only approximately. Nevertheless, the magnitude 
distribution of the irregular satellites can give information 
about the satellite size distribution under the assumption that 
these bodies possess uniform albedos. The cumulative apparent 
magnitude distributions of the satellites of all four giant planets 
are plotted in Figure \ref{cumuplot1}. 
Differences between the cumulative 
satellite counts in the Figure are largely a result of the inverse 
square law. This may be seen in Figure  \ref{cumuplot2}, 
in which the inverse 
square law dependence on distance has been removed 
(Jewitt \& Sheppard 2005, Sheppard et al. 2006).

One discernable conclusion from Figures  \ref{cumuplot1} and 
\ref{cumuplot2} is that the 
cumulative magnitude distributions of the four irregular satellite 
populations have similar slopes. We represent the size distributions 
by power laws, in which the number ofsatellites with radius in the 
range $r$ to $r + dr$ is $n(r) dr = \Gamma r^{-q} dr$,
with $\Gamma$ and $q$ constant. At Jupiter, Saturn, and Uranus, 
the satellite size 
distributions (at assumed constant albedo) are consistent with $q$=2 
(Sheppard \& Jewitt 2003; Kavelaars et al. 2004; 
Jewitt \& Sheppard 2005; Sheppard et al. 2005, 2006). 
These distributions are much flatter than comparable power-law 
representations of the size distributions of the main-belt asteroids 
($q \sim$3.5, with significant size-dependent variations; 
Bottke et al. 2005), small Jovian Trojans 
($q \sim$3.0$\pm$0.3 for radii $<$20 km; Jewitt et al. 2000), or KBOs 
($q$ = 4.0$_{-0.5}^{+0.6}$, Trujillo et al. 2001).
 If the satellites were captured 
from one of these populations, then we infer that the capture 
efficiency was size-dependent, or the satellite size distribution 
has been modified after capture by unspecified processes. 
It should be noted that the Jovian irregulars are imperfectly 
described by a single power law: at radii $<$5 km they follow a 
steeper, $q \sim$3.5, distribution, quite like the classical 
Dohnanyi (1969)  power law. Satellite populations of the other
 planets are less well observed at these small size scales, 
so it is too early to decide whether this steepening of the 
distribution is general.

A second result to be drawn from Figure \ref{cumuplot2} 
 is that, to within 
uncertainties owing to small number statistics, the irregular 
satellite populations of the giant planets are similar. As we 
discuss below, this observation is surprising, given that Jupiter 
and Saturn are gas giants while Uranus and Neptune are ice giants, 
with very different orbit radii, masses, compositions and, 
presumably, formation paths (e.g., Lissauer 2005). Many or most 
of the satellites could be fragments produced collisionally after 
capture. In this case, it would be more reasonable to compare the 
number of satellite dynamical families at each planet. Doing so 
degrades the statistics but takes us to the same conclusion: the 
four very different giant planets possess a handful of irregular 
satellite families.

It is possible, although we think it unlikely, that the observed 
invariance of the irregular satellite populations is a result of 
chance. Different capture mechanisms could operate at different 
planets and just happen to give approximately the same number of 
irregulars (or irregular satellite families) around gas-rich and 
gas-poor planets, with masses spanning the range 17 M$_{\oplus}$
to 310 M$_{\oplus}$  
(Jewitt \& Sheppard 2005). More likely, the satellite invariance 
points to a different capture mechanism, whose efficiency does 
not depend strongly on the details of the planet accumulation 
(hydrodynamic collapse versus ice-rock planetesimal accretion), 
or even on the masses of the planets themselves. The most promising 
mechanism from this perspective is three-body capture, as first 
discussed by Columbo \& Franklin (1971) and explored in more detail 
by Agnor \& Hamilton (2006). Its N-body counterpart may also be 
effective (Astakhov et al. 2003). In these scenarios, the larger 
Hill spheres of the more distant planets (Table \ref{hilltable}) 
help offset their smaller masses.

\section{Case Studies}
In this section we describe three irregular satellites for
which we possess data of unusual quality or quantity. 

\subsection{J VI Himalia}

Prograde Jovian irregular J VI Himalia was discovered 
photographically in 1904 (Perrine 1905). The effective diameter 
of Himalia, determined from combined optical and thermal 
infrared measurements, is about 185 km (Cruikshank et al. 1982). 
It is the dominant member of a family (in $a-i$ space, see
Figure \ref{avsi} and Figure  \ref{aeiplot}) having four secure
members. The others are J VII Elara,
J XI Lysithea, and J XIII Leda. Satellite S/2000 J11 is
potentially also a member but its orbit is poorly established, 
and we here omit it from the list. In Table \ref{himaliafamily} 
 we list diameters
for the other family members based on absolute
magnitudes by Luu (1991) and on the assumption that the 
satellites all have the same ($\sim$3\%) albedo.

The mass of Himalia has been estimated, from its perturbations 
on other satellites (principally J VII Elara), as
4.2 $\pm$ 0.6$\times$10$^{18}$ kg 
(Emelyanov 2005). The prograde family of which Himalia is the 
dominant member has a velocity dispersion significantly larger 
than expected on the basis of numerical models of satellite 
disruption by collision (Nesvorn\'y et al. 2003). Christou (2005) 
explores the possibility that this could be an artifact of 
gravitational scattering of the fragments after disruption using 
models for Himalia mass estimates in the range 
1.7$\times$10$^{18}$ kg to 5.2$\times$10$^{18}$ kg. 
He finds the scattering hypothesis plausible provided 
the mass of Himalia is near the upper end of this range, consistent 
with the estimate based on perturbations by Emelyanov (2005).

The mass of Himalia is apparently known to within $\pm$15\% but the 
volume (and hence the density) is much less accurately determined. 
Images from the Cassini spacecraft at 70$^{\circ}$ phase angle show a
marginally resolved disk (Figure \ref{himalia}), 
with dimensions 150$\pm$20 km 
by 120$\pm$20 km (Porco et al. 2003). Given the limb darkening 
expected at this large phase angle, the larger dimension is 
probably a better approximation to the true size of Himalia, 
as suggested also by the 185 km diameter obtained from
ground-based measurements by Cruikshank et al. (1982). 
In the latter measurement, the accuracy of the diameter is 
limited by uncertainties in the model used to interpret the
thermal flux and is systematic, rather than random, in nature.
These two size estimates give densities of  
$\rho$ = 2400 kg m$^{-3}$ and
 1300 kg m$^{-3}$ for this object. The lower density would suggest 
an ice-rich composition, probably with significant porosity. 
Neither substantial bulk ice nor internal porosity would be
required if the true density is closer to the higher value. 
The factor-of-two difference between the density values is 
probably a meaningful estimate of the systematic uncertainties 
in the determination. In view of this, it seems safe to conclude 
that the composition of Himalia is not significantly constrained
by its estimated density.

The optical reflection spectrum of Himalia is nearly flat, 
but shows a downturn starting at 0.55 $\mu$m that reaches 
its greatest 
depth at about 0.7 $\mu$m (Luu 1991; Jarvis et al. 2000). This band 
has been interpreted as evidence for the presence of hydrated
minerals (Jarvis et al. 2000; Vilas et al. 2006). The near-infrared 
spectrum of J VI Himalia is featureless (Geballe et al. 2002) and 
specifically lacks the 2.0 $\mu$m band due to water. A weak detection 
of a band at 3 $\mu$m (due to water ice or to a hydrated mineral) has 
been claimed (Chamberlain \& Brown 2004) but the data at these 
longer wavelengths have poor signal-to-noise ratios, potentially 
large systematic errors, and their significance is unclear. The 
albedo of Himalia is extraordinarily low: The geometric albedo 
scale in Figure \ref{himalia_sp} shows values of ~3\% across 
the plotted region.
The low albedo is comparable to values measured in the Jovian 
Trojans (Fern{\'a}ndez, Sheppard \& Jewitt 2003) and on the nuclei 
of comets, and suggests (but does not prove) a carbon-rich surface.

\subsection{S IX Phoebe}

The first Saturnian irregular satellite to be discovered, Phoebe 
(Pickering 1899) was also the first to be imaged at high resolution 
from a spacecraft (Porco et al. 2005). The surface of this
107$\pm$1 km 
radius object is heavily cratered (Figure \ref{phoebe}), 
with more than 130 
craters 10 km in diameter or larger (Porco et al. 2005). Craters 
are apparent at all scales down to the (few tens of meters) 
resolution of the best Cassini images. The crater morphology
suggests that most of the features on Phoebe are formed by impact, 
and attest to the long space-exposure of the surface. The largest 
crater is the $\sim$100 km diameter Jason, which is comparable in size 
to Phoebe's radius. With a mean impact speed onto 
Phoebe of $\sim$3.2 km s$^{-1}$ (Zahnle et al. 2003; 
Nesvorn\'y et al. 2003), 
a projectile some 4 km to 5 km in diameter would be needed to create
a 100-km diameter crater (Burchell \& Johnson 2005). The kinetic
energy of such a projectile per unit mass of Phoebe is about 
60 J kg$^{-1}$ (assuming that the projectile and Phoebe have the same 
density). This is about 1\% of the gravitational binding energy 
per unit mass (about 5000 J kg$^{-1}$) of Phoebe, and far short of 
the $\sim$10$^5$ J kg$^{-1}$ needed for catastrophic disruption
of a 107 km radius target (Benz \& Asphaug 1999).

Large impacts like the one responsible for Jason cannot disrupt 
the satellite but must have inflicted substantial damage to the 
interior. As a result, and like many other bodies in the solar 
system, Phoebe is probably internally fractured into a large number 
of competent blocks that are held together by gravity, with void 
spaces in between. The tensile strength of such an assemblage will 
be small. A minimum estimate of the {\it compressive} 
strength is given 
by the $\sim$10 km depth of Jason. This is roughly one-tenth 
the radius 
of the satellite, showing that Phoebe is able to sustain compressive 
stresses about one tenth of the core hydrostatic pressure 
(or $\sim$8 bars) without failure. The overall shape of Phoebe is 
close to a sphere, consistent with a fractured interior in which 
blocks can roll and slip in response to applied stresses. 
However, there is no compelling evidence that Phoebe is a member 
of a satellite family, left behind by an ancient disruptive 
collision. Although Phoebe's orbital inclination is similar to 
those of four other satellites (the others are S/2000 S1, S/2000 S7, 
S/2000 S9, and S/2000 S12; see Figure \ref{avsi} 
and Gladman et al. 2001), 
its other orbital elements do not appear to be clustered 
(Figure \ref{avse}), 
giving no evidence for a related dynamical family of impact-ejected 
fragments.

Phoebe's dark surface (the mean visual geometric albedo is 
0.081$\pm$0.002, with spatial variations of a factor of two; 
Simonelli et al. 1999) may not be representative of the bulk 
interior. Cassini images show several types of evidence for 
stratigraphic layering on Phoebe. First, layering is directly 
exposed in the walls of some craters (Figure \ref{layer}), 
with the top 
layer being the darkest. Second, some small craters appear bright 
relative to their surroundings, suggesting that bright material 
has been excavated by these impacts from beneath a darker surface 
layer. Third, down-slope motion is apparent from vertically 
aligned streaks in the walls of various craters 
(e.g., see A and B in Figure \ref{slump}). Material appears to have 
fallen from the walls, exposing bright (more ice rich?) material. 
Slumped material is evident beneath the crater walls 
(see C in Figure \ref{slump}), 
showing the importance of down-slope motion 
even though the surface gravity is only  
$\sim$0.05 m s$^{-2}$. Together, 
these observations suggest that Phoebe's surface has been darkened, 
perhaps by the loss of volatiles or some other form of 
space-weathering, relative to the brighter, more pristine material 
underneath.

Spatially resolved infrared spectra of the surface of Phoebe were 
taken by the Cassini spacecraft (Clark et al. 2005). They reveal 
(Figure \ref{phoebe_spec})
 a number of distinct bands associated with water ice 
(1.5 $\mu$m, 2.02  $\mu$m and 2.95  $\mu$m), trapped
CO$_2$ (4.26  $\mu$m), 
probable CN (2.42  $\mu$m and 4.5  $\mu$m),
 and weaker bands due to other 
compounds, including probable phyllosilicates and organics. 
A broad feature near 1.0 $\mu$m
may be caused by electronic transitions
in a mineral containing Fe$^{2+}$. The low albedo of Phoebe is 
attributed to surface organics, perhaps processed by interaction 
with charged particles from the solar wind and cosmic rays.
The water ice bands are less deep in the interiors of some 
craters than on bright surfaces outside the rim-walls. This might 
indicate that the volatiles on Phoebe have an external origin,
perhaps resulting from the impact of comets and the subsequent 
freezing of cometary matter as a thin veneer on the satellite.

The mass of Phoebe has been measured from gravitational deflections 
on passing Voyager and Cassini spacecraft. Combined with the 
measured dimensions, the mass indicates a bulk density for Phoebe 
of 1630$\pm$45 kg m$^{-3}$ 
(Porco et al. 2005). This is too dense to match 
a pure ice composition and too underdense to match pure rock,
unless the bulk porosity is a very high 40\% or more. Most likely,
Phoebe is a composite of ices and rock (consistent with surface 
spectroscopy) with an uncertain but nonzero porous fraction. 
Porosity is an expected consequence of energetic collisions that 
have internally fragmented Phoebe. Its survival is possible 
because of the low core hydrostatic pressure, 
$P_c \sim 4 \pi/3 G \rho^2 R^2$, 
with $G$ = 6.67$\times$10$^{-11}$ N kg$^{-2}$ m$^2$ for 
the Gravitational constant. 
Substituting, we estimate
$P_c \sim$ 8$\times$10$^6$ N m$^{-2}$, or only 80 bars.

The bulk density has been used by Johnson \& Lunine (2005) to 
argue that Phoebe is a captured KBO. They note that the 
mass-weighted mean density of the regular Saturnian satellites 
Mimas, Enceladus, Tethys, Dione, Rhea, and Iapetus is 
$\sim$1300$\pm$130 kg m$^{-3}$.
They assert that Phoebe is significantly denser, 
being more comparable to Pluto and Triton (both of which have 
uncompressed densities $\sim$1900 kg m$^{-3}$). 
They further invoke a 
compositional model and calculate that the measured density is 
consistent with the known solar abundances of the elements and 
a protoplanetary nebula in which most of the carbon is locked 
up in CO (as opposed to CH$_4$, which is likely to dominate in the 
dense, hot subnebulae of the planets). Although interesting,
these considerations are not compelling both because there is 
no simple relation between density and formation location, and 
because the relation between density and object size is not a 
simple correlation. For example, the high densities of Pluto and 
Triton are not matched by other KBOs: (20000) Varuna has  
$\rho \sim$ 1000 kg m$^{-3}$ 
(Jewitt \& Sheppard 2002; Takahashi \& Ip 2004), 
2001 QG298 has $\rho$ = 600 to 1000 kg m$^{-3}$
(Sheppard \& Jewitt 2004;
Takahashi \& Ip 2004), and (47171) 1999 TC36 has
$\rho$ = 550 to 800 kg m$^{-3}$ 
(Stansberry et al. 2005). It is amusing to note that 
the {\it low} density of Jovian Trojan (617) Patroclus has been used 
to argue that this object, too, must be from the Kuiper belt 
(Marchis et al. 2006). The argument is similar in spirit to 
the one advanced for Phoebe, but opposite in relative density!
The connection between the bulk density and the formation location
remains obscure.

\subsection{N I Triton}

Triton is by far the largest satellite likely to have an origin 
by capture. Key parameters include its diameter (2706$\pm$2 km), 
density (2061$\pm$7 kg m$^{-3}$), semimajor axis of its orbit around 
Neptune (354800 km; 14.4 Neptune radii, and about 0.003$r_H$), 
eccentricity (0.00002), and retrograde orbit with an inclination 
of 156.8$^{\circ}$. 
Three scenarios have been proposed for capture of this 
object: energy dissipation through tidal friction, gas drag, 
and three-body interactions including collisions. All three 
scenarios infringe on the fantastic: Triton crystallizes the 
problems that surround the capture of all irregular satellites.

Tides exerted between Neptune and Triton lead to torques and
internal dissipation of energy that could act to shrink and 
circularize the satellite orbit and also cause a modest 
evolution in the inclination (McKinnon \& Leith 1995). In this 
scenario Triton would enter Neptune's Hill sphere from a probable 
source location in the Kuiper belt, and tidal dissipation would 
convert the orbit from a temporarily captured retrograde one 
into permanent capture. Triton is much more dissipative than 
Neptune and so the dissipated orbital energy would appear as heat 
inside Triton, with potentially profound consequences for the 
thermal evolution and surface geology of this body 
(Figure \ref{triton}). 
The tiny eccentricity of Triton's current orbit provides 
compelling evidence for the action of tides but it is not obvious 
that tidal dissipation is responsible for capture itself. 
McKinnon \& Leith (1995) argue that Triton is too far from Neptune 
for tidal dissipation to act on the timescale of a temporary 
capture. Either the satellite was not captured through tidal 
dissipation, or its current orbit results from modification
by other processes after tidal damping.

Gas-drag capture in an extended, collapsing envelope, as proposed 
for the gas-giant planets Jupiter and Saturn (Pollack et al. 1979), 
seems very unlikely at Neptune (or Uranus). The latter planets are 
relatively gas-free, with distinctly nonsolar compositions dominated 
by the metals C, N, and O. The ice giants never experienced a phase 
of hydrodynamic collapse and so offer little opportunity for 
satellite capture in this way. However, it is possible that Uranus 
and Neptune were attended by equatorial gas and dust disks at the 
late stages of their accretion. At Neptune, there is no strong 
evidence for such a disk. Neptune lacks a system of substantial 
regular satellites that might indicate disk accretion but, if such 
a system ever existed it would probably have been disrupted by the 
capture of Triton. Indeed, the absence of a substantial system of 
regular satellites at Neptune has been advanced as evidence for 
Triton's origin by capture (Goldreich et al. 1989). Ice giant 
Uranus does possess regular satellites (Ariel, Umbriel, Titania, 
Oberon, and Miranda) in the 500 km to 1600 km diameter range that 
could have formed through accretion in an equatorial disk. These 
satellites have been used to estimate a (very high) reconstituted 
satellite disk surface density 
$\sigma \sim$ 3.4$\times$10$^7$(r/R$_U$)$^{-1.5}$ kg m$^{-2}$, 
where $r/R_U$ is the radial distance in units of Uranus' radius 
(McKinnon \& Leith 1995). The same researchers then showed that 
Triton, if moving on a grazing (retrograde) orbit passing through 
a similar disk at Neptune, would experience  non-negligible 
drag forces that could lead to capture. Problems with this scenario 
include the short lifetime of the disk to viscous spreading 
(perhaps as little as 1000 years): How likely could it be that 
one of the largest objects in the Kuiper belt would encounter the 
dense protosatellite disk at exactly the right time to be 
captured? More seriously, very dense protosatellite disks appear 
incompatible with evidence from the satellites themselves 
(e.g., Callisto should have formed so rapidly in such a disk 
that captured gravitational binding energy should have led to 
whole-body differentiation, whereas moment-of-inertia data 
show only partial differentiation). Perhaps the mass flowed 
through the disk toward the planet, and was not all present at 
one time (Canup \& Ward 2002, 2006). Lastly, the regular 
satellites of Uranus might have formed by an entirely different 
process, such as accretion from debris blown out from the planet 
following a massive impact (Stevenson et al. 1986). In this case, 
Uranus would have no relevance to what might have happened at 
Neptune.

Three-body interactions might have captured Triton. In the most 
extreme three-body interaction, a collision within the Hill sphere 
between Triton and a pre-existing regular satellite of sufficient
mass could have stabilized the orbit and destroyed the regular 
satellite system simultaneously (Goldreich et al. 1989). Relative 
to capture by gas drag, the collisional hypothesis has a much 
longer timescale for action (since it is not limited by the survival 
of a hypothesized protosatellite disk) but a much lower
probability of occurring. The latter is given roughly by the 
ratio of the cross-section of Triton to the area of its orbit 
and is $\sim$10$^{-5}$, for an unbound body passing once 
through the Neptune 
system. Alternatively, Triton could have entered the Neptune Hill 
sphere as a binary, been tidally split from its companion by Neptune 
and then captured, with the excess energy carried away by the 
escaping secondary (Agnor \& Hamilton 2006).

\section{Dynamics and Collisions}

The numbers and orbital distributions of the irregular satellites 
reflect both the details of the capture process and subsequent 
dynamical and collisional evolution. Early models of the 
satellites focussed on their long-term dynamical stability. 
As our observational assessments of the irregular satellites 
have improved, the additional importance of collisional and 
other destructive processes is becoming clear. The emerging 
view is that the modern-day irregular satellites are survivors 
from initial populations that were at least a few times, and 
perhaps orders of magnitude, larger than now. Both dynamical 
and collisional losses may have been important.

The large semimajor axes (a few hundred planetary radii) of 
irregular satellites, along with their highly inclined and 
eccentric orbits, make them susceptible to external perturbations 
from the Sun and other planets. These perturbations are stronger 
at apoapse distances, and are the source of some of the 
interesting dynamical features of these objects. For instance, 
as shown by Henon (1970), Jupiter¿s retrograde irregulars are
more stable than their prograde counterparts, a dynamical
feature that is consistent with the observed overabundance 
of former objects.

The long-term stability of an irregular satellite is 
affected by its orbital eccentricity and inclination 
(Hamilton \& Burns 1991). 
In general, orbital stability is defined as the nonexistence 
of secular changes in the semimajor axis of an object.
The variations of the orbital inclination and eccentricity 
at this state are assumed to be negligibly small. In case of 
irregular satellites, however, these variations, combined 
with the perturbative effect of the Sun, play a significant 
role in the general dynamics of these objects. For instance, 
the solar perturbation that is the primary cause of the 
precessions of the orbital planes of irregular satellites 
affects the motion of Jovian irregulars approximately four 
times more than the motion of the Moon around Earth. Solar 
tugs create the exchange of angular momentum between an 
irregular satellite and the Sun, and as shown by Kozai (1962), 
enlarge the orbital eccentricity to high values 
at large inclinations. For the system of Jovian irregulars 
this happens within a timescale of approximately 180 years 
for prograde satellites and 65 years for the retrograde ones 
(Carruba et al. 2002).

The absence of irregular satellites at inclinations
$55\leq i \leq 130$ deg. (Figure \ref{avsi})  
is a likely result of planetary and solar perturbations 
driving the periapses of irregular satellites to small values 
by increasing their orbital eccentricities through the 
above-mentioned mechanism, known as the Kozai resonance 
(Carruba et al. 2002, Nesvorn\'y et al. 2003). At this state, 
the longitude of periapse, $\omega_p$, and the orbital eccentricity,
$e_p$, of the satellite vary as functions of its orbital inclination,
$i_p$, as (Innanen et al. 1997)

\begin{equation}
{\sin^2}{\omega_p}=\,0.4\,{\csc^2}{i_p},
\label{omega}
\end{equation}

\begin{equation}
{(e_p^2)_{\rm max}}={1\over 6}\,\Bigl[1-5\cos (2{i_p})\Bigr].
\label{eccen}
\end{equation}

\noindent
As $e_p$ cannot be less than zero, Equation  \ref{eccen}  
shows that the 
Kozai resonance may occur for orbital inclinations in the 
range of 39.2$^\circ \le i \le 140.8^{\circ}$, 
roughly coinciding with the observed 
absence of highly inclined irregular satellites 
(Carruba et al. 2002; Figure \ref{avsi}).

The stability limits of prograde and retrograde irregular 
satellites are asymmetric. That is, retrograde irregulars 
are stable on larger orbits. As shown by Hamilton \& Krivov 
(1997), the three-body interaction 
between a prograde satellite, its host planet, and the Sun 
can be the cause of this effect. Numerical simulations by 
Nesvorn\'y et al. (2003) suggest that this asymmetry may have
roots in the precession of the orbit of the irregular satellite, 
and may have been caused by the evection resonance 
(Touma \& Wisdom 1998, Nesvorn\'y et al. 2003). In this resonance, 
the period of the precession of the apoapse of the satellite's 
orbit becomes equal to the period of the planet around the Sun. 
Solar tides on the satellite, particularly at apoapse, cause 
its apocenter to drift outward. Once close to the Hill radius, 
the satellite becomes unstable and escapes the system, leading 
to the selective depletion of prograde irregulars.

Irregular satellites of all inclinations are dynamically unstable
when on highly eccentric orbits. These objects may collide with 
the central planet or other regular satellites, or, more usually,
may leave the planet's Hill sphere. The probability of collision
per orbit, $P$, for an irregular satellite with a periapse distance
inside the orbit of a prograde satellite with a physical radius 
of $r_G$ and an orbital radius of $R_G$, is approximately given 
by $P\simeq {({r_G}/2{R_G})^2}$. 
This expression yields a value equal to $5 \times {10^{-7}}$ 
for collision with, for instance, Callisto 
(${r_G}=$2400 km, ${R_G}\simeq 26{R_J}$). A Callisto-crossing 
irregular satellite with an orbital period of one year will 
survive for only $\sim$10$^6$ yr. 
For this reason, it is not surprising 
that Jovian irregular satellites avoid Galileans completely 
(the smallest perijove belongs to J XVIII and is approximately 
80 Jupiter radii).

Irregular satellites could also collide with external objects. 
Observed groups of irregulars with similar orbits imply that 
previous collisions might have occurred between a parent body 
and a fast-moving impactor. The possibility of an impact between 
an irregular satellite and a comet, or an escaped Trojan or 
asteroid, in the present state of the solar system, is small 
(Nakamura \& Yoshikawa 1995, Zhanle et al. 2003). However, such 
collisions might have been important in the past when small 
bodies were more abundant in the outer solar system.
Collisions might also occur among irregular satellites. Initial 
estimates of the collisional timescales (Kessler 1981) have been 
superseded by numerical simulations in which our recently 
improved knowledge of the satellite populations has been taken 
into account (Nesvorn\'y et al. 2003). Figure \ref{nes} indicates the 
possible importance of collisions in model satellite systems 
integrated over 4.5 Gyr (Nesvorn\'y et al. 2003). For each of 
four large irregular satellites of the giant planets, the figure 
shows the number of collisions with a counter-rotating swarm of 
test satellites, as a function of the semimajor axes of these 
satellites. The eccentricities and inclinations of the test swarm 
were set to be typical of the known irregulars at each planet. 
Figure \ref{nes} shows that, at each planet, there is a local maximum 
in the collision probability close to the orbit of the target 
satellite ({\it arrows} mark the semimajor axes of these satellites). 
In addition, there is a general trend towards larger numbers 
of collisions at smaller semimajor axes, resulting from the  
dependence of the Keplerian orbital periods.

Satellite-satellite collisions would occur at speeds of several 
km s$^{-1}$, generally resulting in the destruction of the small 
impacting satellites and the creation of impact craters on the 
larger bodies. For example, Figure \ref{nes} suggests that retrograde
satellites of Jupiter with orbits near Himalia's would have 
significant likelihood of collision in the age of the Solar 
system, perhaps explaining the paucity of such satellites 
(cf. Figure \ref{avsi}). Jupiter's known retrograde irregulars orbit 
at larger distances where they are immune to destructive 
sweeping by Himalia and other prograde satellites. A more 
striking result is seen in Figure \ref{nes}
for Neptune's Nereid. 
This large, prograde irregular (the diameter is 340$\pm$50 km;
Thomas et al. 1991) has a large cross-section for sweeping up
retrograde satellites on comparably small orbits. Neptune¿s 
known irregulars (other than massive Triton) are indeed located 
at larger distances, far beyond Nereid's reach (Figure \ref{avsi}).

Sufficiently energetic impacts can result in the breakup of the 
target object and the creation of satellite dynamical families. 
Indeed, satellite clustering has long been recognized as evidence 
for the past break-up of precursor satellites (Kuiper 1956; 
Pollack et al. 1979). As in the asteroid belt, much of the mass 
of the disrupted satellite should reaccrete under its own gravity 
into a rubble-pile type object, perhaps containing large void 
spaces and having small tensile strength. Dominant family members 
like Himalia and Ananke around Jupiter could well be objects that
have reaccreted after shattering collisions. In the modern Solar 
system, projectiles large enough to shatter 100-km 
scale bodies are very rare, 
and it is safe to associate these events with a much earlier 
(but postcapture) epoch when the density of projectiles would 
have been much higher than now (Nesvorn\'y et al. 2004). After 
collision, a small fraction of the target satellite mass would 
escape immediate fall-back, creating the dynamical family. A 
key clue as to the correctness of this picture is that the 
velocity dispersions within families are comparable to the 
gravitational escape speeds of the largest family members.
For example, the Carme and Ananke families at Jupiter have 
velocity differences 5 $\le \delta V \le$ 50 m s$^{-1}$ and 
15 $\le \delta V \le$ 80 m s$^{-1}$, respectively 
(Nesvorn\'y et al. 2003, 2004). The escape velocities from Carme 
($\sim$46-km diameter) and Ananke ($\sim$28-km diameter) are about 
25 m s$^{-1}$ and 15 m s$^{-1}$, respectively, assuming bulk densities 
$\sim$2000 kg $^{-3}$. 
Another indication is provided by high resolution 
images of Saturn's Phoebe (Figure \ref{phoebe}), where the 
$\sim$100-km diameter of the Jason crater is comparable to 
the radius of the 
satellite. A slightly larger impact would have disrupted 
the satellite.

Possible evidence for the collisional erosion of the irregular 
satellites has been produced by dust detectors on the Galileo 
spacecraft (Krivov et al. 2002). Micron-sized dust grains in 
both prograde and retrograde orbits in the 50 R$_J$ to 300 R$_J$ 
radius range are consistent with erosion rates expected from 
bombardment by interstellar and interplanetary dust. The dust 
number density of $\sim$10 km$^{-3}$, 
while extraordinarily low, is about 
10 times the dust density in the local interstellar medium.

Mauna Kea survey observations (Sheppard \& Jewitt 2003) of the 
Jupiter system show that no irregular satellites exist with 
semimajor axes between the outermost Galilean satellite, 
Callisto (at 26 R$_J$), and the innermost irregular satellite, 
Themisto (semimajor axis 101 R$_J$). Numerical simulations by
Haghighipour show that 
the Galilean satellites are capable of destabilizing objects in 
this region. This is shown in Figure \ref{sim1}, where, for values of 
eccentricity larger than 0.2, and for inclinations beyond 20$^\circ$, 
the region between Callisto and Themisto is naturally unstable. 
As the eccentricities and inclinations of particles increase, 
their orbits become unstable to perturbations by the two outer
Galilean satellites of Jupiter, Ganymede, and Callisto. About 
three-fourths of the unstable objects are ejected from the 
Jupiter system and the remainder are destroyed by impacting 
(primarily) the planet.

Some of the irregular satellite orbits exist in secular 
resonance with each other. These resonant orbits can reveal 
details of the dynamics, origin, and evolution of their 
corresponding bodies. The transition time from a non- or 
near-resonant state to a resonance may take between $10^7$  
years for a non-Kozai resonance, to $10^9$ years for the 
Kozai resonance. Saha \& Tremaine (1993) suggested that the 
former is reached through the evolution of a satellite's 
orbit subject to some dissipative force, whereas the latter 
indicates that Kozai resonant orbits may be primordial 
implying that the Kozai resonance did not play an important 
role in capturing irregular satellites since  not many such 
resonant satellites have been discovered. The resonances 
among irregular satellites are rare (only 8 retrograde 
satellites among all currently known irregulars have resonant 
orbits, cf. Nesvorn\'y et al. 2003), and can only be found 
among retrograde objects.

\section{Origin of Irregular Satellites}

It is very unlikely that irregular satellites were formed by 
accretion in a circumplanetary disk, as were the regular satellites 
(Canup \& Ward 2002, 2006). Neither the inclination distribution
nor the large sizes of the orbits of the irregular satellites 
can be reconciled with an origin in a circumplanetary disk. 
Instead, these objects must have been formed elsewhere and 
later been captured into their current orbits around their host 
planets. Numerical simulations of planetary growth indicate 
that most planetesimals in the vicinity of the growing planets 
were scattered out of the planetary region of the Solar system. 
[A small (1\% to 10\%) fraction of these bodies were emplaced in 
the Oort cloud but most were launched into interstellar space
 and are forever lost. There are no efficient dynamical pathways 
from the Oort cloud to the irregular satellites and so we consider 
these objects no further.] The irregular satellites could be 
objects (``asteroids'' or ``comets'') from nearby heliocentric orbits 
that happened to escape dynamical ejection during the planet 
growth phase. Alternatively, the irregular satellites might have
been captured from source regions in the Kuiper belt. 
In some models, gravitational interactions with migrating giant 
planets clear substantial mass (perhaps several tens of
M$_{\oplus}$) from 
the young Kuiper belt (Morbidelli et al. 2005; Tsiganis et al. 2005), 
raising the possibility that the irregular satellites could be 
captured KBOs.

Three basic mechanisms have been suggested to account for the 
formation of irregular satellites:

(1) Capture due to the sudden mass-growth of Jupiter, 
the so-called pull-down mechanism (Heppenheimer \& Porco 1977);

(2 )Permanent capture through dissipation due to gas drag 
(Pollack et al. 1979; Astakhov et al. 2003; Cuk \& Burns 2004); and

(3) Capture through three-body interactions 
(Columbo \& Franklin 1971). 
In the following we discuss these mechanisms in detail.

\subsection{Pull-Down Capture}

The formation of the giant planets of our Solar system has been 
the subject of intense study. Jupiter and Saturn are gas giants,
with most of their masses contained in hydrogen and helium that
must have been acquired directly from the Solar nebula. 
Arguments persist about the precise mechanism of the formation 
of these objects. The widely accepted core accretion model
suggests that a solid body, consisting of high molecular weight 
material (metals), grew through binary accretion from the 
protoplanetary disk in much the same way as the terrestrial 
planets are thought to have formed through the collision of 
kilometer-sized objects. Materials in the cores of giant planets 
include the same refractory substances (silicates, organics) 
as in the terrestrial planets with the addition of simple ices, 
notably water, that carry about 50\% of their condensible mass. 
According to this scenario, the growth of the core continued up 
to a critical mass, generally estimated as near 10 $M_{\oplus}$  
(the escape velocity from the core is then of order 20 km s$^{-1}$), 
whereupon the core underwent a runaway growth and attracted 
its adjacent nebular gas through a hydrodynamic flow.

The most widely studied problem with the traditional core accretion 
model is that the core must form fast enough to reach its critical 
mass before the nebular gas dissipates (Pollack et al. 1996). 
Direct observations of gas disks in other systems are difficult, 
but measurements of thermal radiation from dust disks around 
solar mass stars (e.g., Carpenter et al. 2005) suggest that the 
timescale for disk dissipation is  $\sim$10 Myr. 
Erratic dust production, 
possibly owing to collisions between large bodies, decays on 
timescales ten times longer (see Rieke et al. 2005). Until recently, 
the estimated core growth times have been longer than the 
inferred disk decay times, making the acquisition of a massive 
gaseous envelope impossible. An alternative scenario, namely the 
disk instability model (Mayer et al. 2002), 
avoids this timescale problem by forming the core in just a few 
thousand years. In this model, the protoplanetary disk is locally 
dense enough to collapse spontaneously under its own gravity 
without need for a central core to grow first. However, this 
mechanism suffers from difficulties in losing heat on timescales 
short enough to cool the nebula sufficiently to trigger its 
collapse down to planetary dimensions before the solids are 
dispersed by differential rotation in the disk.

Whether by the core accretion mechanism, or through the disk 
instability scenario, the key feature of gas-giant formation 
is a runaway growth in mass, most of it gaseous hydrogen and 
helium. As suggested by Heppenheimer \& Porco (1977), a sudden 
increase in a planet¿s mass would cause a jump in its Hill 
radius, trapping temporary satellites of the growing planet into
permanently bound retrograde orbits. Pull-down capture allows 
small bodies in the neighborhood of the Lagrangian points of a 
growing gas-giant planet (i.e., in a 1:1 mean-motion resonance 
with the latter object) to be captured in stable orbits, provided 
at the time of their capture, they are moving in the Hill sphere
of the growing planet with a low relative velocity 
(Heppenheimer \& Porco 1977; Vieira Neto et al. 2004). 
This mechanism also requires the timescale of the increase of 
the planetary mass to be small compared to the time that the
object spends in the planet's Hill sphere.

Recently, it has been shown that the pull-down mechanism can also 
account for the permanent capture of prograde irregular satellites. 
By backwards integrating the equations of motion of a restricted 
three-body system (Sun-Jupiter-Satellite), and allowing the mass 
of Jupiter to decrease, Vieira Neto et al. (2006) have simulated 
the dynamics of an already captured prograde irregular satellite 
and obtained a limit of instability beyond which the satellite 
would escape the system. Given the time-reversibility of dynamical 
systems, their results indicate that pull-down capture can also 
occur for prograde objects. The process in this case is more 
complicated than the capture of retrograde satellites and occurs 
in two steps. For a growing Jupiter, an irregular satellite at 
approximately 0.85 Hill Radii, and in the vicinity of the $L_1$ or 
$L_2$ Lagrangian points, enters a region of temporary capture where 
it is locked in an evection resonance (Saha \& Tremaine 1993). The 
semimajor axis of the satellite in this region undergoes 
oscillations. If the satellite continues its inward migration and 
passes the stability boundary at 0.45 Hill Radii, it will be 
captured in a permanent prograde orbit. The irregular satellites 
Leda, Himalia, Lysithea, and Elara may have been captured through 
this mechanism (Vieira Neto et al. 2006).

The pull-down mechanism may not be able to explain the origin of 
the irregular satellites of Uranus or Neptune, because these 
ice-giant planets grew slowly with little or no runaway growth 
in mass due to capture of nebular gas. In the case of Jupiter, 
for instance, as shown by Vieira Neto et al. (2004), a sudden 
increase of at least 10\% in Jupiter's mass is needed in order 
for its retrograde irregular satellites to be captured in stable 
orbits.

\subsection{Gas-Drag Capture}

The runaway growth in the mass of the gas giants offers another 
way to trap satellites. Young and still-forming Jovian-type planets 
initially possess bloated envelopes, hundreds of times larger than 
the resulting planets, which shrink as they cool by radiation into 
space. Solid bodies passing through these gaseous envelopes will 
slow down owing to frictional dissipation by gas drag. In some 
cases, gas drag could cause solid bodies moving on initially 
heliocentric orbits to become bound to the planets. This is the 
essence of the gas-drag capture mechanism, first explicated by 
Pollack et al. (1979).

In gas-drag capture, the irregular satellites are thought to be 
passing asteroids or comets whose orbits became temporarily 
captured about the planets and then converted to bound orbits 
by frictional losses. Capture efficiency is a function of size: 
Small bodies would burn up or spiral into the central planet in 
a short time, whereas large bodies would scarcely feel the effects
of drag and could not be retained. Complexity (and uncertainty) 
in the gas-drag model arises because the bloated envelope is itself
a dynamic, short-lived structure. The sudden collapse of the
envelope permits objects spiraling toward destruction to escape
their fate, but also ends further opportunities for capture.
Later collisions among captured satellites can change their 
shapes and size-distribution. In a recent paper, by considering 
an accretion disk (Lubow et al. 1999; d'Angelo et al. 2002; 
Bate et al. 2003) instead of an extended atmosphere, 
Cuk \& Burns (2004) 
have argued that gas-drag retardation can indeed account for 
the capture of the prograde (Himalia) cluster of Jovian 
irregular satellites. We merely comment that such a model 
is necessarily based on a large number of weakly constrained 
and uncertain parameters, particularly relating to the geometry,
density, and time-dependence of the in-flowing circumplanetary gas.

Two consequences of the gas-drag scenario are the implication of a 
minimum mass for irregular satellites for which an observational 
assessment is yet to be made, and lower values of orbital 
eccentricity for smaller irregulars. Although there is some 
evidence of higher eccentricity for larger irregular satellites, 
such evidence is statistically insignificant. In any case, 
postcapture collisional modification of the orbits might conceal 
any trends produced during gas-drag capture. There is one piece 
of observational evidence compatible with the past action of 
gas drag. As explained in the previous section, the orbits of 
several satellites occupy weak resonances: dissipation by drag 
from residual gas could explain how the satellites fell into 
such resonant states (Saha \& Tremaine 1993; Whipple \& Shelus 1993).

\subsection{Three and N-Body Interactions}

The observation that the four giant planets have similar numbers 
of irregular satellites, measured down to a common size, does not 
sit easily with the gas-drag hypothesis for capture 
(Jewitt \& Sheppard 2005). Only Jupiter and Saturn are gas giants
 with massive hydrogen and helium envelopes needed for capture 
(Pollack et al. 1996). Uranus and Neptune are comparatively 
gas-free ice giants, with only $\sim$1 M$_{\oplus}$ of H$_2$ 
and He compared with $\sim$300 and $\sim$100 M$_{\oplus}$ 
in Jupiter and Saturn, respectively. While it is 
conceivable that residual gas at Uranus and Neptune might have 
helped capture irregular satellites there, the observed approximate 
invariance of the irregular satellite populations among planets
with very different compositions, structures, masses and modes 
of formation, is certainly not a natural consequence of the
gas-drag hypothesis.

Likewise, the pull-down capture hypothesis is viable, if anywhere, 
only about the gas-giant planets. Only they experienced the runaway 
growth in mass needed to expand the Hill spheres on a sufficiently 
rapid timescale. The ice-giant planets in contrast grew by the
steady accretion of ice-rock planetesimals and were never able 
to attain a runaway configuration, which is why they are deficient 
in gas. The mere existence of irregular satellites around the ice 
giants argues against pull-down (and gas drag) as likely agents 
of capture.

The existence of the satellite dynamical families proves that the 
satellites have been subject to collisions with other bodies since 
the time of their capture. It is a small step from this observation 
to the conjecture that physical collisions or scattering interactions
between small bodies could have led to the capture of the satellites 
to begin with. Interactions within the planetary Hill sphere can 
lead to the excess kinetic energy being converted to other forms 
(heat or comminution energy) if there is a physical collision, or 
simply being carried away by one of the bodies after a close 
encounter (Columbo \& Franklin 1971; Weidenschilling 2002). 

As a variant on three-body interactions, a wide binary object 
could be split following an approach to a massive planet, with 
one component becoming bound and the other being ejected, 
carrying with it the excess energy from the system 
(Agnor \& Hamilton 2006). Because a considerable fraction of the
KBOs are thought to be binaries (perhaps 10\% or more; 
Stephens \& Noll 2006), the supply of these objects might be 
large enough to account for the irregular satellite populations.

Capture of quasi-satellites may be another way to form irregular 
satellites. Quasi-satellites are bodies in 1:1 co-orbital resonance
with the planets. Kortenkamp (2005) has argued that 5\% to 20\% of 
planetesimals scattered by a planet will become quasi-satellites, 
and he showed that a significant fraction of these objects pass 
through the planetary Hill sphere at low relative velocities. 
This makes the capture of these objects easy provided there is 
some form of dissipation. For example, energy loss by gas-drag 
in the solar nebula can lead to the capture of quasi-satellites 
without the need for circumplanetary gas drag. The mass-growth 
of the planet can have a similar effect. However, Kortenkamp's 
simulations show that quasi-satellite formation is efficient only 
when the orbital eccentricities are enlarged to values 
($\sim$0.1 or more) much greater than now possessed by the planets.

Although proposed more than three decades ago, three-body and 
N-body capture models have received little attention until recently, 
perhaps because the densities of the involved objects are small, 
and their assumed dynamical interaction times are correspondingly 
long compared to the age of the Solar system. The key is to realize 
that the density of these objects at the epoch of capture may have 
been vastly higher than in the modern-day Solar system. Despite the 
difficulty in the applicability of the three-body interaction 
scenario to Neptunian irregulars (the latter objects might have 
been destroyed or scattered from and throughout the system as a 
result of interaction with Triton and Nereid (Cuk \& Gladman 2005), 
the biggest advantage of this scenario over the others 
is its independence from the mechanism of the formation of giant 
planets in our Solar system.

\subsection{Source Regions}
  
The source regions from which the irregular satellites were derived 
remain unknown. However, it is possible to divide these sources 
into local and nonlocal. Source regions local to the host planets 
are favored in terms of capture efficiency because they are likely 
to provide low velocity encounters with a smaller energy barrier 
to capture objects in permanently bound orbits. These local source
regions include those planetesimals that were originally moving 
in the vicinity of the growing planets but were neither scattered 
away nor absorbed by collision with the planets. If the sources 
were local to the planets, then the irregular satellites assume
new significance as survivors from the long-gone population of 
bodies that collided to build the high molecular-weight cores of 
the planets.

Nonlocal source regions are those that feed objects into the 
Hill spheres of the planets from remote locations within the 
protoplanetary disk. Encounters with objects from distant 
sources tend to occur at higher mean velocities and permanent 
capture occurs with reduced but nonzero efficiency. For example, 
it has been argued that the Trojan asteroids of Jupiter could 
have been captured chaotically from a Kuiper belt source in a 
late-stage clearing event in the Solar system 
(Morbidelli et al. 2005). This event is predicated on the assumed
crossing of the 2:1 mean motion resonance between Jupiter and 
Saturn, itself driven by torques acting on a long-lived particle 
disk (proto-Kuiper belt) of assumed mass 
30 M$_{\oplus}$ to 50 M$_{\oplus}$ (Tsiganis et al. 2005).

Observationally, it might be possible to distinguish locally 
derived satellites from nonlocal ones. If irregular satellites 
were captured from the Kuiper belt, for instance, then some of 
their observable properties might resemble similar properties 
of the KBOs. The comparison is presently very difficult, in part
because the parameters of many irregular satellites remain poorly 
known. Furthermore, the mean size of the well-studied KBOs 
(few $\times$100 km to 2500 km diameter) is substantially greater 
than the mean size of the well-studied Trojan asteroids 
(few $\times$10 km to 100 km), so that size-dependent gradients 
in the measured properties are of potential concern. 
The better-determined physical properties of the Jovian irregular 
satellites are compared with those of Jupiter's Trojans, 
and with the KBOs, in Table \ref{comparisontable}. 
A reasonable conclusion to draw 
from the comparisons made in this table is that the irregular 
satellites do not physically resemble the KBOs, apparently 
contradicting the hypothesis that the irregular satellites are 
captured KBOs (Morbidelli et al. 2005). However, several 
evolutionary effects must be considered before this conclusion 
can considered firm.

\section{Epilogue}

Examples of irregular satellites have been known for more than a 
century, while their significance as captured objects has been
recognized for at least half this time. Still, many of the most 
basic questions about these objects remain unanswered. 
The mechanism of capture is not known [we possess several 
(quite different) ideas, any or all of which could be wrong]. 
The source region, from which the irregular satellites were 
derived, has yet to be identified. Neither do we know when the 
satellites were captured, although we can be sure that capture 
was not recent. Nevertheless, it is hard to deny that our 
understanding of the irregular satellites is steadily improving, 
particularly in their role as probes of early conditions in the 
Solar system. The systematics of the satellite populations are 
beginning to be revealed by powerful ground-based survey 
observations. We know that irregular satellites are abundant 
around all four giant planets, that they are predominantly 
retrograde, and that they are confined to the central 50\% of 
their planet¿s Hill spheres. Many belong to dynamically related 
families probably resulting from postcapture collisions. 
Irregular satellites are almost certainly survivors from larger 
initial satellite populations that have been depleted through 
collisional and dynamical losses. Saturn's irregular satellite 
Phoebe has been closely examined, showing a heavily cratered 
surface coated with dirt, spectral traces of water, and other 
ices that suggest, to some, an origin in the Kuiper belt. 
Eventually, we will need in situ measurements from spacecraft to 
better measure the compositions. In the mean time, advances on 
the irregular satellites are expected from continued, even deeper 
surveys, and from detailed physical observations using the 
largest telescopes.

\acknowledgments

This work was supported by a NASA Planetary Astronomy grant to D.J. 
N.H. is supported by the NASA Astrobiology Institute under 
cooperative agreement NNA04CC08A at the University of Hawaii.

\clearpage
\begin{figure}[h]
\epsscale{0.9}
\begin{center}
\plotone{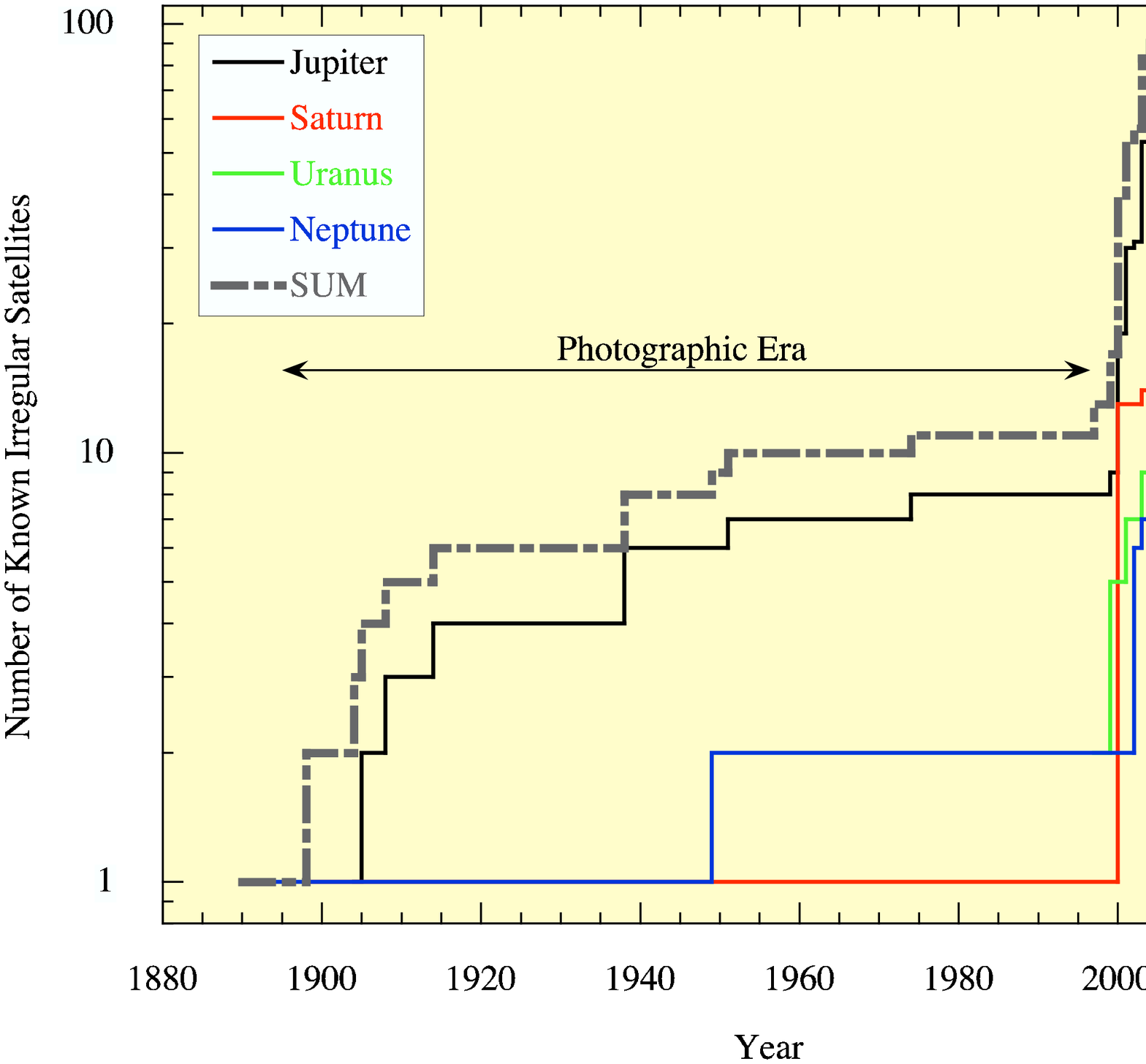}
\caption{Number of known irregular satellites of the giant 
planets (Jupiter, black; Saturn, red; Uranus, green; and 
Neptune, blue) as a function of date. The sum of these 
populations is also shown (gray dash-dot line). The sudden 
jump in the known satellite populations at the start of the 
21st century is the result of the application of large-format 
CCD surveys.
\label{nvst}} 
\end{center} 
\end{figure}

\clearpage
\begin{figure}[h]
\epsscale{0.9}
\begin{center}
\plotone{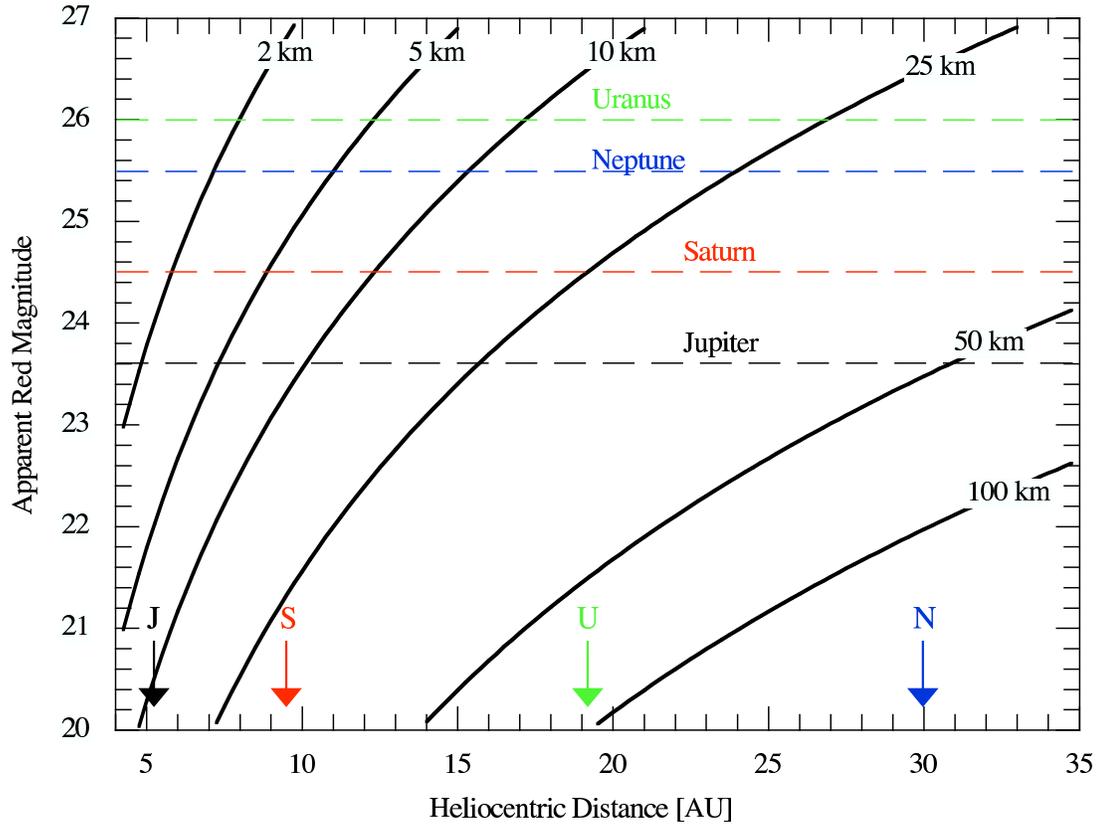}
\caption{Diameters of objects viewed in scattered light as 
a function of their heliocentric distance and apparent red 
magnitude. A red geometric albedo of 0.04 has been assumed. 
Dashed horizontal lines show, for each planet, the approximate 
magnitude limits to which published satellite surveys are 
complete. (Figure adapted from Sheppard et al. 2006) 
\label{sizes}} 
\end{center} 
\end{figure}

\clearpage
\begin{figure}[h]
\epsscale{0.8}
\begin{center}
\plotone{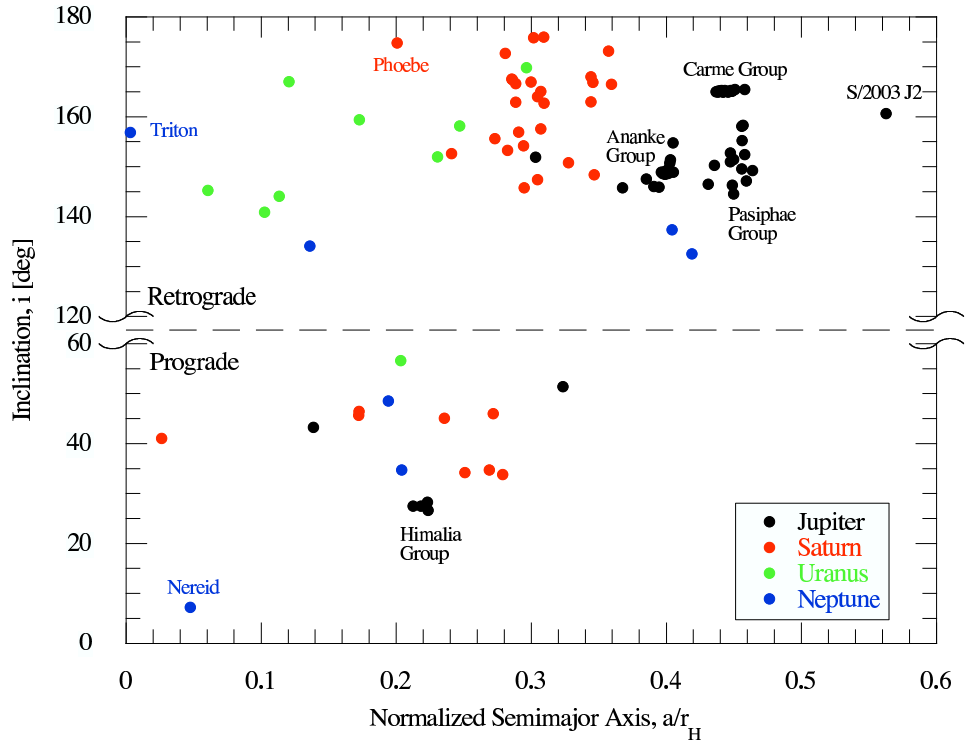}
\caption{Semimajor axis scaled to the Hill sphere radius 
versus orbital inclination, for the irregular
satellites of the giant planets known as of November 1, 2006.  
The region 60$^{\circ} \le i \le$ 120$^{\circ}$
contains no satellites and is not plotted. 
\label{avsi}} 
\end{center} 
\end{figure}

\clearpage
\begin{figure}[h]
\epsscale{0.8}
\begin{center}
\plotone{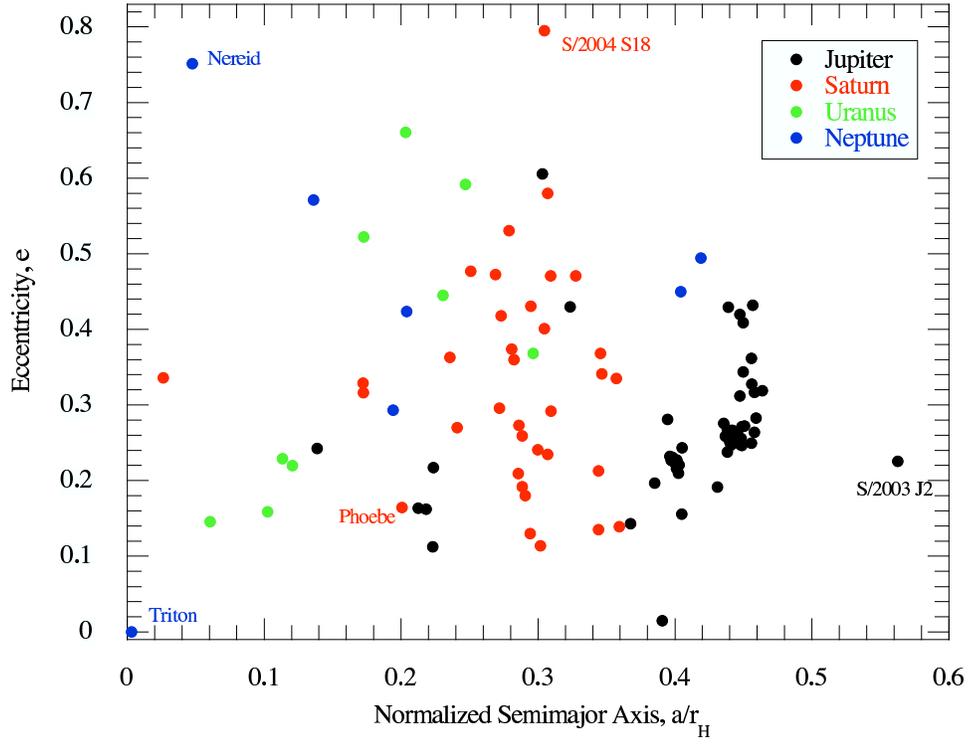}
\caption{Semimajor axis scaled to the Hill sphere radius 
versus orbital eccentricity, for the irregular
satellites of the giant planets known as of November 1, 2006. 
\label{avse}} 
\end{center} 
\end{figure}

\clearpage
\begin{figure}[h]
\epsscale{0.8}
\plotone{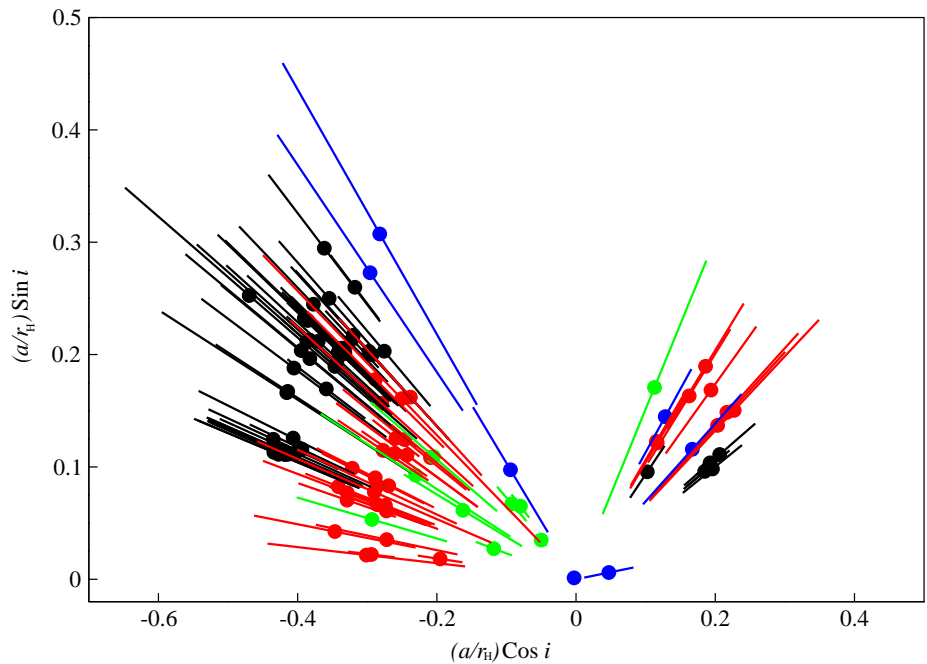}
\caption{Alternative plots showing the distribution of irregular satellites 
at Jupiter (black), Saturn (red), Uranus (green), and Neptune (blue).  
The plot shows ($a$/$r_H$)$cos(i)$ versus ($a$/$r_H$)$sin(i)$,
where ($a$/$r_H$) is the semimajor axis in units of the Hill radius, 
and $i$ is the orbital inclination. The distance of each satellite from
the origin gives the semimajor axis, the angle from the $x$-axis gives the 
inclination (prograde objects plot with $x >$0) and the radial excursion 
from periapse to apoapse is indicated by the length of the line. 
\label{aeiplot}}
\end{figure}

\clearpage
\begin{figure}[h]
\epsscale{0.8}
\begin{center}
\plotone{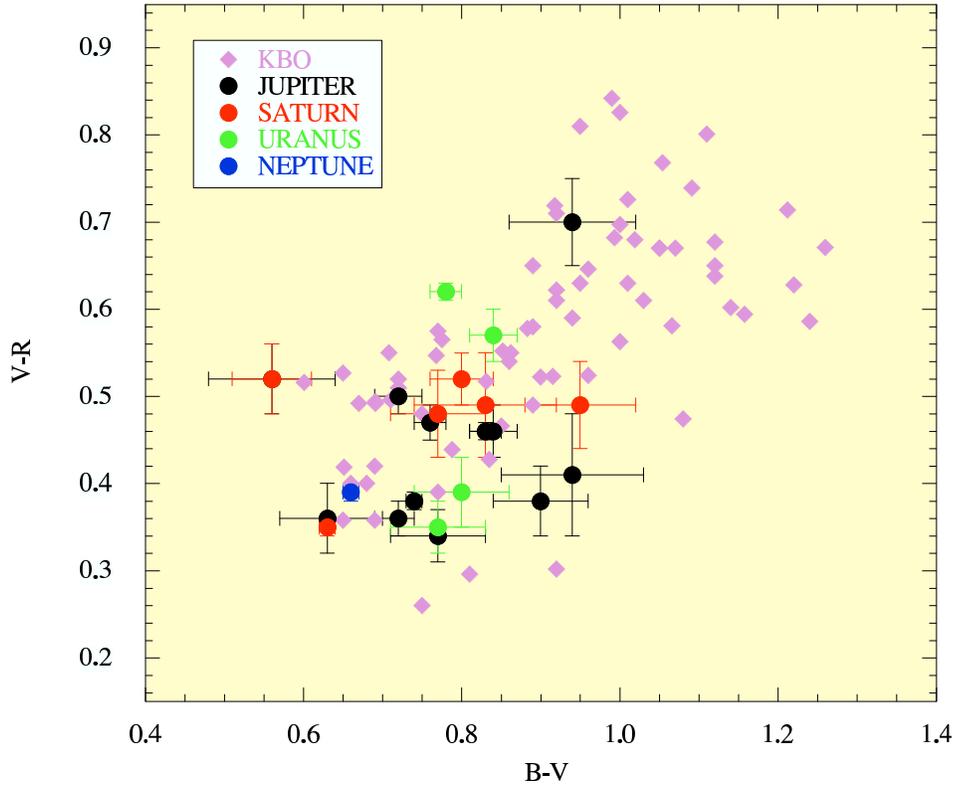}
\caption{Color-color plane for irregular satellites of 
Jupiter (black), Saturn (red), Uranus (green) and
Neptune (blue) compared
with the colors of Kuiper Belt Objects (purple diamonds).   
Only satellites with color uncertainties
1$\sigma \le$ 0.1 mag are plotted.  Satellite data from 
Grav et al 2003.  KBO data are from
Doressoundiram et al. 2002, Boehnhardt et al. 2002 and 
unpublished measurements by
the authors. 
\label{color_color}} 
\end{center} 
\end{figure}

\clearpage
\begin{figure}[h]
\epsscale{0.8}
\begin{center}
\plotone{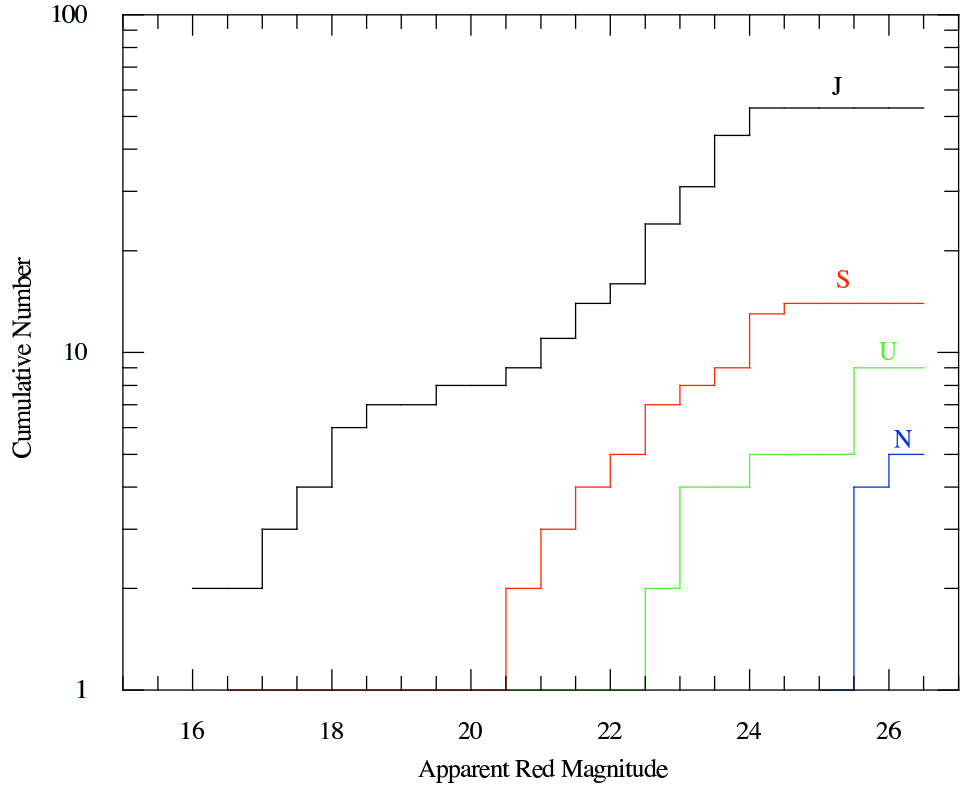}
\caption{Cumulative distributions of the apparent red magnitudes 
of the irregular satellites of the planets. 
Figure from Jewitt \& Sheppard (2005). 
\label{cumuplot1}} 
\end{center} 
\end{figure}

\clearpage
\begin{figure}[h]
\epsscale{0.8}
\begin{center}
\plotone{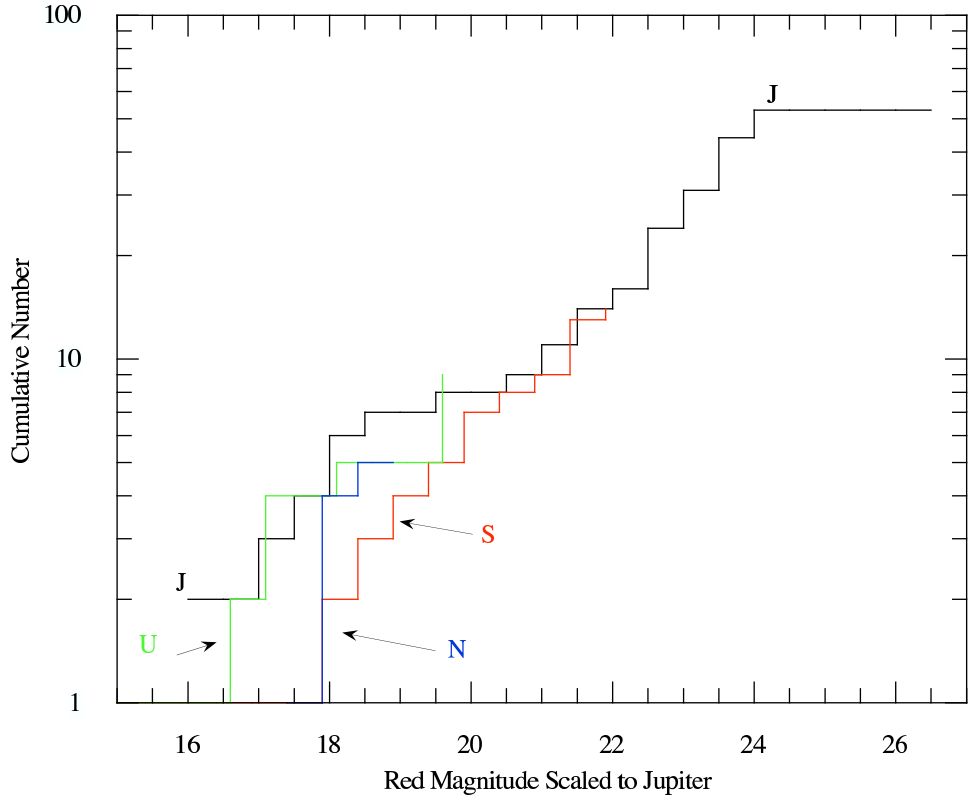}
\caption{Cumulative distributions of the magnitudes of the 
irregular satellites of the planets corrected to Jupiter's 
opposition distance by the inverse square law. 
Figure from Jewitt \& Sheppard (2005). 
\label{cumuplot2}} 
\end{center} 
\end{figure}

\clearpage
\begin{figure}[h]
\epsscale{0.8}
\begin{center}
\plotone{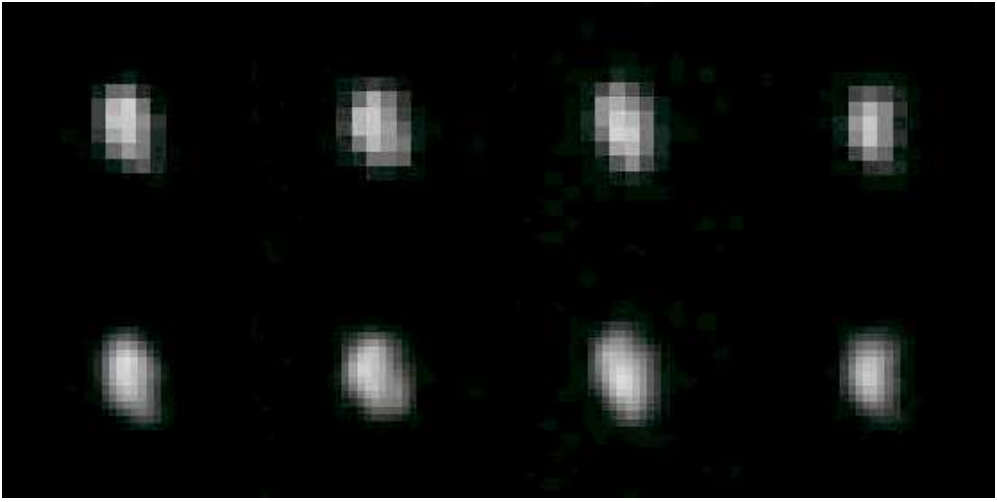}
\caption{Images of J VI Himalia from the Cassini spacecraft.  
Images in the top row show Himalia
at four different times in a $\sim$4.5 period.  
Smoothed versions of these images are shown in the bottom row.
From Porco et al. (2003). 
\label{himalia}} 
\end{center} 
\end{figure}

\clearpage
\begin{figure}[h]
\epsscale{0.6}
\begin{center}
\plotone{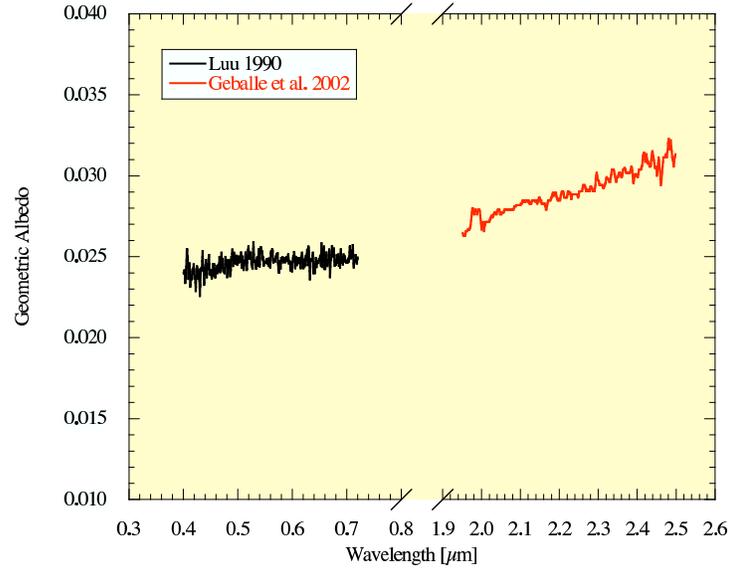}
\caption{Composite optical to near-infrared reflection spectrum
 of J VI Himalia. The optical spectrum from Luu (1991) has
been normalized by eye to the reflection spectrum in the 2.0 to
2.5 $\mu$m wavelength range by Geballe et al. (2002).
No useful data exist in the 0.7 to 2.0 $\mu$m spectral range.
\label{himalia_sp}} 
\end{center} 
\end{figure}

\clearpage
\begin{figure}
\epsscale{0.8}
\begin{center}
\plotone{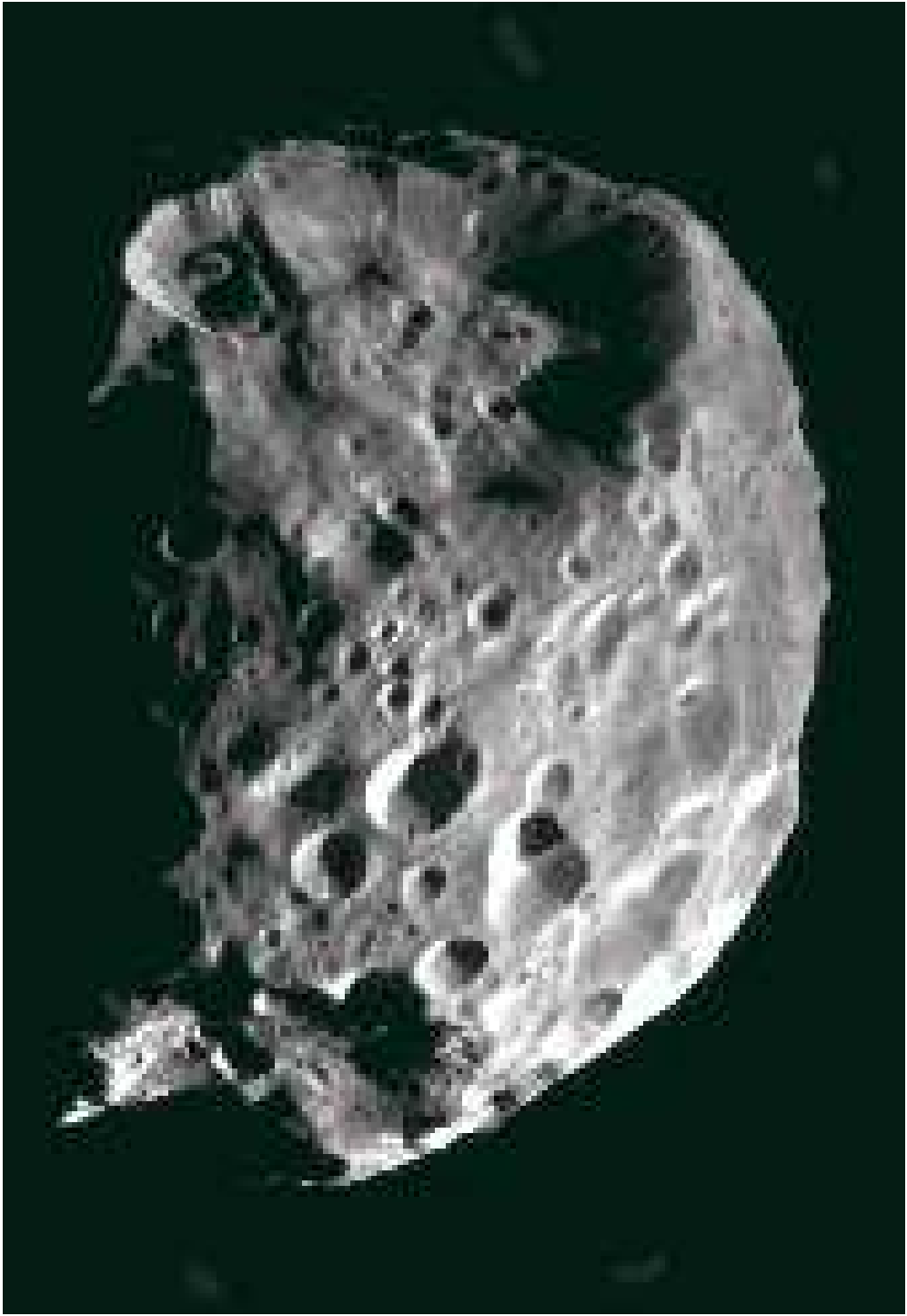}
\caption{Image of Phoebe recorded from the Cassini spacecraft on 
June 11, 2004.  The phase
angle in this image is 84$^{\circ}$ and the image scale approximately 
200 meters per pixel.   
Image from Porco et al. 2005 and courtesy Cassini Imaging Team 
and NASA/JPL/Space Science Institute. 
\label{phoebe}} 
\end{center} 
\end{figure}

\clearpage
\begin{figure}[h]
\epsscale{0.5}
\begin{center}
\plotone{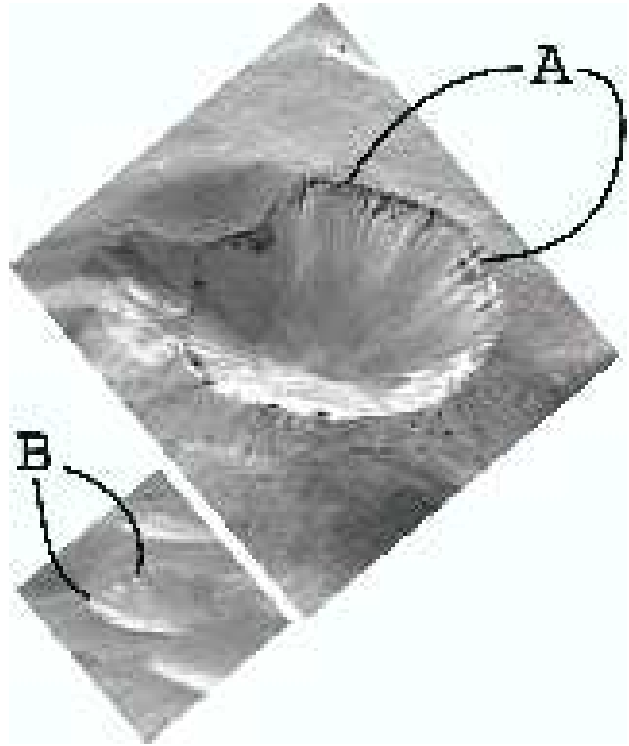}
\caption{Layering in the walls of two craters on Phoebe, 
indicated by letters A and B.  The large
crater, Euphemus, is about 20 km in diameter, the smaller 
(nameless) about 8 km.
Image courtesy Cassini Imaging Team and NASA/JPL/Space 
Science Institute. 
\label{layer}} 
\end{center} 
\end{figure}

\clearpage
\begin{figure}[h]
\epsscale{0.6}
\begin{center}
\plotone{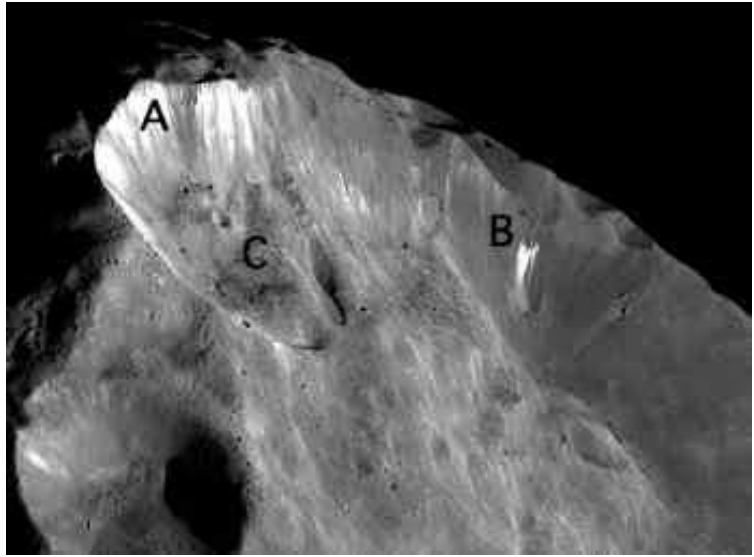}
\caption{Close-up showing material slumped down the wall 
of a large crater
on Phoebe, apparently exposing clean ice.   
Image courtesy Cassini Imaging Team and NASA/JPL/Space 
Science Institute. 
\label{slump}} 
\end{center} 
\end{figure}

\clearpage
\begin{figure}[h]
\epsscale{0.9}
\begin{center}
\plotone{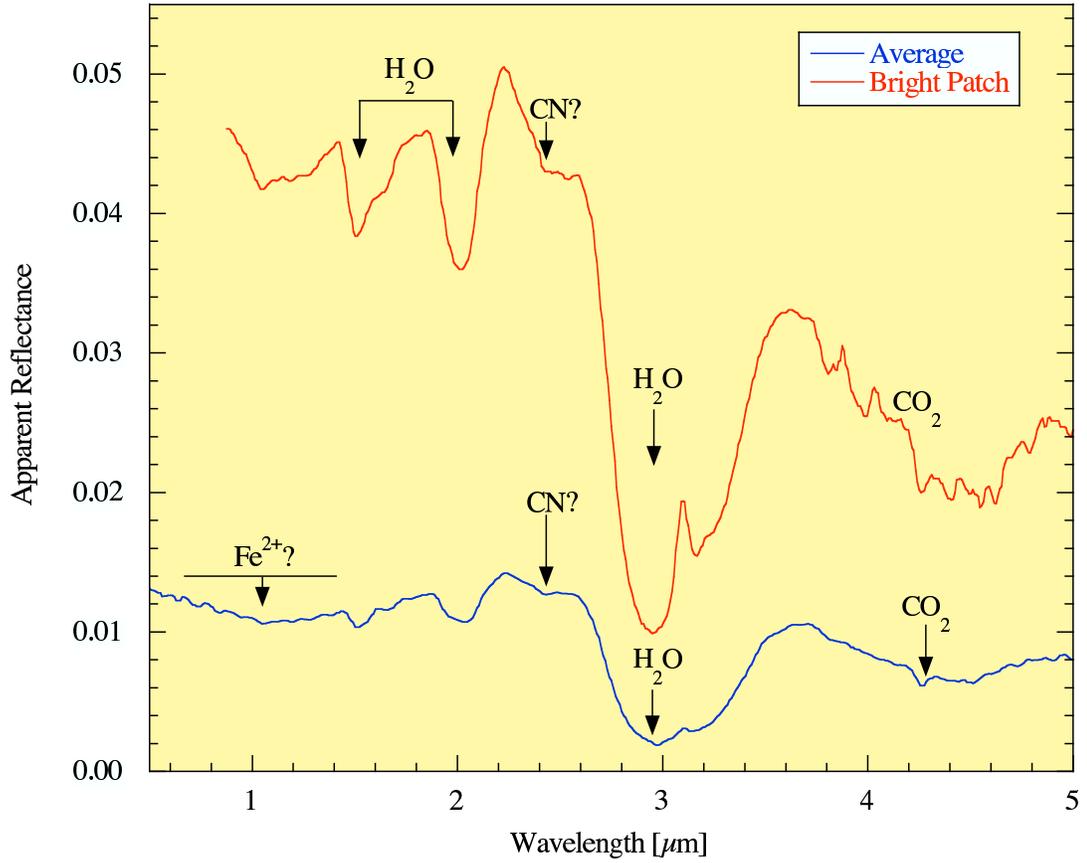}
\caption{Spectra of Phoebe from the Cassini Visible and Infrared 
Mapping Spectrometer. Red and
blue curves show spectra of a bright (icy) patch on the surface 
and a global average.
Adapted from Clark et al. 2005
\label{phoebe_spec}} 
\end{center} 
\end{figure}

\clearpage
\begin{figure}[h]
\epsscale{0.6}
\begin{center}
\plotone{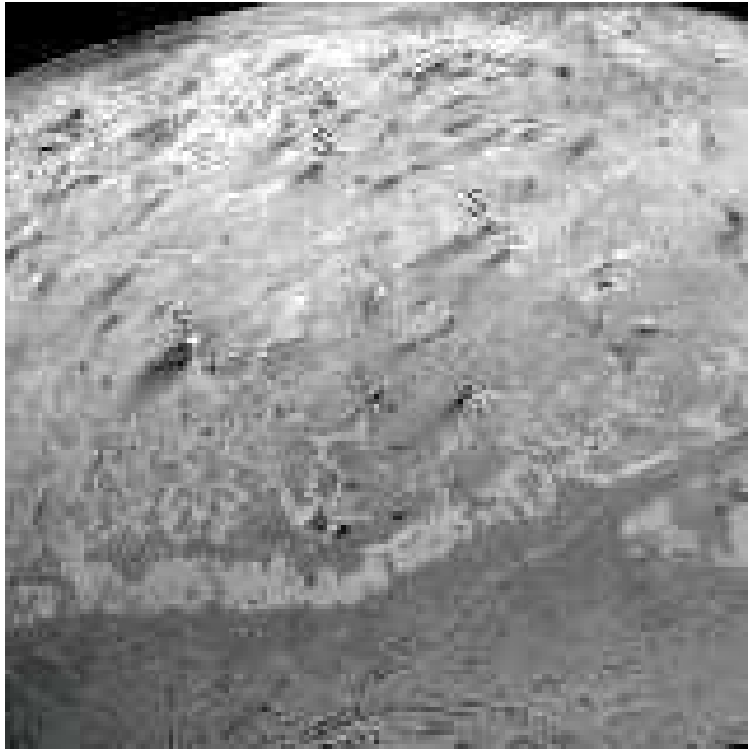}
\caption{South polar region of Neptune's giant retrograde 
satellite Triton as imaged by the Voyager 2 spacecraft.  This image 
shows a relatively crater-free (young) ice surface and is 
divided into two parts. At the top is the south polar
region, across which are deposited dark streaks (marked S).  
These may be caused by vented plumes of material that
is carried by winds across the surface. At the bottom are 
smooth plains cut by a double trench-like lineament. Only 
a few, small craters are evident. Region shown is about 
800 km wide. Image courtesy NASA. 
\label{triton}} 
\end{center} 
\end{figure}

\clearpage
\begin{figure}[h]
\epsscale{0.7}
\begin{center}
\plotone{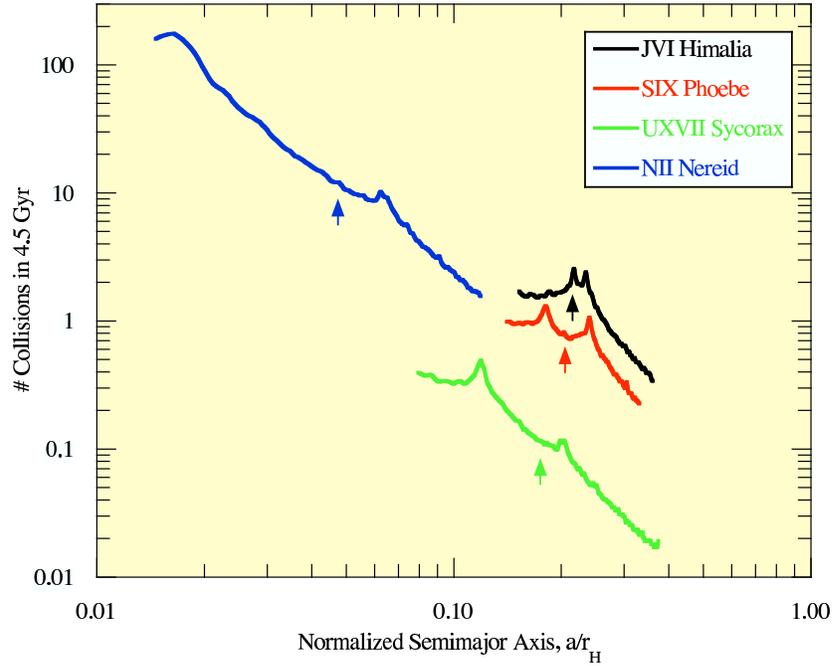}
\caption{Number of collisions between selected large irregular 
satellites and
test satellites experienced in 4.5 Gyr 
as a function of the semimajor axis measured
in units of the Hill sphere radius.  The curves for each of
four large irregular satellites mark the 
radial excursions of these bodies in units of the appropriate
Hill sphere radius.  
The test satellites were assumed to orbit in 
a direction opposite to the large irregular satellites with 
eccentricities and inclinations
typical of the real irregulars at each planet.   The
semimajor axes of the large irregulars are marked with arrows.  
Figure adapted
from Nesvorn\'y et al. (2003).  
\label{nes}} 
\end{center} 
\end{figure}

\clearpage
\begin{figure}[h]
\epsscale{0.7}
\begin{center}
\vskip -0.18in
\plotone{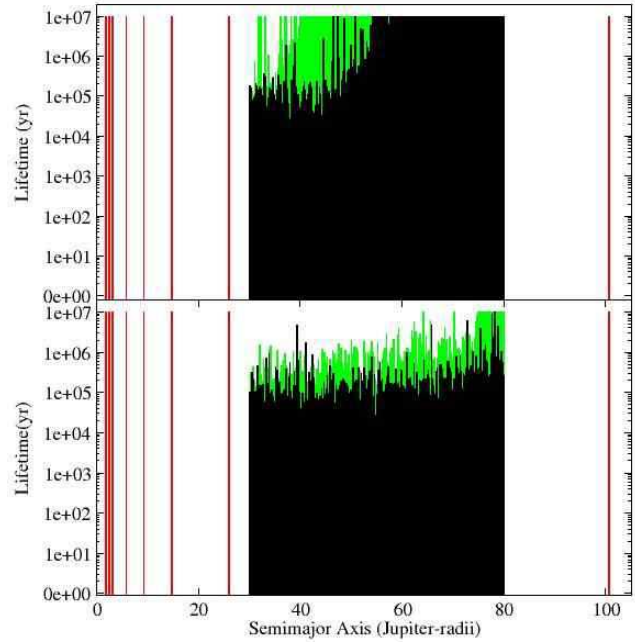}
\caption{Lifetimes of hypothetical irregular satellites 
of Jupiter computed
in the region from 30 to 80 Jupiter-radii. In the top graph, irregulars 
in black have zero initial orbital inclinations, and their initial orbital
eccentricities are equal to 0.2. The objects in green in the top graph 
depict irregular satellites with initial orbital inclinations of 
20$^\circ$, and eccentricities of 0.4. In the lower graph, 
the orbital inclination of black objects is 60$^\circ$, and those
of the green ones are 120$^\circ$. The orbital eccentricities of all
particles in the lower graph are 0.6. Vertical red lines mark the semimajor
axes of known satellites (Galileans and other regular 
satellites at $a_p \le$ 26 R$_J$,
Themisto at $a_p$ = 102 R$_J$).
\label{sim1}} 
\end{center} 
\end{figure}

\clearpage
\begin{deluxetable}{lcccccc}
\tabletypesize{\scriptsize}
\tablecaption{Hill Spheres of the Giant Planets \label{hilltable}}
\tablewidth{0pt}
\tablehead{
\colhead{Planet} & \colhead{$m_p$\tablenotemark{a}} & 
\colhead{$a_p$ [AU]\tablenotemark{b}}   & 
\colhead{$r_H$ [AU]\tablenotemark{c}}   & 
\colhead{$r_H$ [deg]\tablenotemark{d}}  & 
\colhead{$\Delta m$\tablenotemark{e}}   & 
\colhead{$N_i$\tablenotemark{f}}}
\startdata
Jupiter   & 310  &  5 & 0.35 & 4.8 &  0 & 55 \\
Saturn   & 95  &  10 & 0.43 & 2.8 &  2.6 & 14 \\
Uranus   & 15  &  20 & 0.47 & 1.4 &  5.9 & 9 \\
Neptune   & 17  &  30 & 0.77 & 1.5 &  7.6 & 7\tablenotemark{g}  \\
\enddata
\tablenotetext{a}{Planet mass in units of Earth's mass 
($M_{\oplus}$ = 6$\times$10$^{24}$ kg).}
\tablenotetext{b}{Semimajor axis in AU}
\tablenotetext{c}{Radius of Hill sphere in AU}
\tablenotetext{d}{Projected angular radius of Hill 
sphere in degrees at opposition}
\tablenotetext{e}{Magnitude decrement 
$\Delta m$ = 5log$_{10}$[$a(a-1)$/($a_J$($a_J$-1)], 
where $a_J$ is the Sun-Jupiter distance} 
\tablenotetext{f}{Total number of reported irregular satellites}
\tablenotetext{g}{Including Triton}
\end{deluxetable}

\clearpage
\begin{deluxetable}{lcccl}
\tabletypesize{\scriptsize}
\tablecaption{Giant Planet Satellite Counts \label{moonsurveys}}
\tablewidth{0pt}
\tablehead{
\colhead{Planet} & \colhead{$N_r$\tablenotemark{a}} & 
\colhead{$N_i(pro)$\tablenotemark{b}}   & 
\colhead{$N_i(ret)$\tablenotemark{c}}  & 
\colhead{$\Sigma N$\tablenotemark{d}}     }
\startdata
Jupiter   & 8  &  6 & 49 & 63   \\
Saturn     &  21 & 8 & 27 & 56  \\
Uranus   & 18  &  1 & 8 & 27  \\
Neptune   & 6  &  4 & 4  & 13 \\	
SUM        & 53 & 19 & 88 & 159 \\	

\enddata

\tablenotetext{a}{Number of regular satellites}
\tablenotetext{b}{Number of prograde ($i <$90$^{\circ}$) 
irregular satellites}
\tablenotetext{c}{Number of retrograde ($i >$90$^{\circ}$) 
irregular satellites}
\tablenotetext{d}{Total number of satellites}
\end{deluxetable}

\clearpage
\begin{deluxetable}{lccccl}
\tabletypesize{\scriptsize}
\tablecaption{Published Irregular Satellite Surveys 
\label{surveys}}
\tablewidth{0pt}
\tablehead{
\colhead{Planet} & \colhead{$m_R$\tablenotemark{a}} & 
\colhead{$A$\tablenotemark{b}}   & 
\colhead{$N$\tablenotemark{c}}  & 
\colhead{Facility\tablenotemark{d}} & 
\colhead{Reference}     }
\startdata

Mars      & 23.5 & 3.0 & 0 & CFHT 3.6-m & Sheppard et al. (2004) \\
Jupiter   & 21.5  &  12 & 1 & UH 2.2-m & Sheppard and Jewitt (2003)  \\
Jupiter   & 22.5  &  4.4 & 9 & UH 2.2-m & Sheppard and Jewitt (2003)  \\
Jupiter   & 23.2  &  12.4 & 10 & CFHT 3.6-m & Sheppard and Jewitt (2003)  \\
Jupiter   & 22.5  &  6.7 & 1 & CFHT 3.6-m & Sheppard and Jewitt (2003)  \\
Saturn     &  22.0 & 1.3 & 3 & ESO 2.2 & Gladman et al. (2001)  \\
Saturn     &  24.5 & 3.0 & 8 & CFHT 3.6-m & Gladman et al. (2001)  \\
Saturn     &  22.0 & 7.0 & 1 & Hopkins 1.2-m  & Gladman et al. (2001)  \\
Saturn     & 26+   & 3+ & 22 & Subaru 8-m & unpublished \\
Uranus   & 23.5  &  0.08 & 2 & Palomar 5-m & Gladman et al. (1998)  \\
Uranus   & $\sim$25  & 1.1 & 4 & CFHT 3.6-m, CTIO 4-m &  Kavelaars et al. 
(2004) \\
Uranus   & 26.1  &  3.5 & 2 & Subaru 8-m & Sheppard et al. (2005)  \\
Neptune   & 25.5  &  1.4 & 5 & CFHT 3.6-m, CTIO 4-m & Holman et al. (2004)  \\
Neptune   & 25.8  &  1.75 & 1 & Subaru 8-m & Sheppard et al. (2006)  \\
\enddata

\tablenotetext{a}{Limiting red magnitude of the survey}
\tablenotetext{b}{Area surveyed in square degrees.  
In cases where the survey area is not explicitly reported, 
we have estimated this quantity to the best of our ability 
from the data provided.}
\tablenotetext{c}{Number of new satellites reported}
\tablenotetext{d}{Telescope employed 
(CTIO = Cerro Tololo InterAmerican Observatory,
UH = University of Hawaii, CFHT = Canada France 
Hawaii Telescope 3.6-m)}
\end{deluxetable}

\clearpage
\begin{deluxetable}{lccccl}
\tabletypesize{\scriptsize}
\tablecaption{The Himalia Family \label{himaliafamily}}
\tablewidth{0pt}
\tablehead{
\colhead{Satellite} & 
\colhead{$a/R_J$\tablenotemark{a}} & 
\colhead{$e$\tablenotemark{b}}   & 
\colhead{$i$\tablenotemark{c}}  & 
\colhead{$m_R(1,1,0)$\tablenotemark{d}}  & 
\colhead{$D_e$\tablenotemark{e}}}
\startdata
J VI Himalia   & 160.5  &  0.162 & 27.5 & 7.60$\pm$0.03 &  185 \\
J VII Elara     &  164.4 & 0.217 & 26.6 & 9.44$\pm$0.02  & 79 \\
J XI Lysithea   & 164.1  &  0.112 & 28.3 & 10.65$\pm$0.03 & 45  \\
J XIII Leda   & 156.4  & 0.164 & 27.5  & 12.56$\pm$0.10 &  19 \\
\enddata

\tablenotetext{a}{Orbital semimajor axis, expressed in units of 
Jupiter's radius, taken to be $R_J$ = 71,400 km.}
\tablenotetext{b}{Orbital eccentricity}
\tablenotetext{c}{Orbital inclination in degrees 
(relative to the local Laplace plane)}
\tablenotetext{d}{Absolute red magnitude from Luu (1991)}
\tablenotetext{e}{Estimated effective diameter in km}
\end{deluxetable}

\clearpage
\begin{deluxetable}{lclll}
\tabletypesize{\scriptsize}
\tablecaption{Comparison of Properties\label{comparisontable}}
\tablewidth{0pt}
\tablehead{
\colhead{Quantity} & \colhead{Symbol} &
\colhead{Irregulars} & 
\colhead{Jovian Trojans} & 
\colhead{KBOs}     }
\startdata

Geometric Albedo\tablenotemark{a} & $p_v$   &  
0.04$\pm$0.01 &  0.041$\pm$0.002  & 0.10$\pm$0.05  \\
Size Distribution Index\tablenotemark{b} & $q$              
 & 2.0$\pm$0.5 & 3.0$\pm$0.3 & 4.0$\pm$0.3 \\
Largest Example [km] & $D_{max}$ & 370$\times$195 (Hektor)
 & 150 - 185  & 2400 \\
Mean Spectral Gradient [\%/1000\AA] \tablenotemark{c}& 
$\overline{S'}$  & 6$\pm$4 & 10$\pm$1  & 23$\pm$2\\
Min, Max Spectral Gradient [\%/1000\AA] \tablenotemark{c} & 
$S_{min}', S_{max}'$  & -5, 20 & 3, 25  & 2, 40\\
Binary Fraction [\%] \tablenotemark{d} & $f_B$   & ?  &  1?  & 
11$_{-2}^{+5}$  \\
\enddata

\tablenotetext{a}{Irregulars: Cruikshank et al. 1982, 
Jovian Trojans: Fernandez et al 2003,
Kuiper belt: Cruikshank et al. 2006 (average of 7 objects 
observed at thermal wavelengths from space, diameters 
100 km to 600 km).}
\tablenotetext{b}{Irregulars: Sheppard and Jewitt 2003, 
Jewitt and Sheppard 2005; Jovian Trojans: 
Jewitt et al. 2000, KBOs: Trujillo et al. 2001}
\tablenotetext{c}{Irregulars: Grav and Bauer 2007 
(Saturn satellites only); Jovian Trojans: Jewitt 2002;
 KBOs: Jewitt 2002}
\tablenotetext{d}{Irregulars: No data; Jovian Trojans: 
F. Marchis, personal communication, KBOs: Stephens and Noll 2006}
\end{deluxetable}


\clearpage

\begin{thebibliography}{}

\bibitem[Agnor \& Hamilton(2006)]{2006Natur.441..192A} Agnor, C.~B., \& 
Hamilton, D.~P.\ 2006, \nat, 441, 192 - 194.

\bibitem[Astakhov et al.(2003)]{2003Natur.423..264A} Astakhov, S.~A., 
Burbanks, A.~D., Wiggins, S., \& Farrelly, D.\ 2003, \nat, 423, 264 

\bibitem[]{} Bate, M. R., Lubow, S. H., Ogilvie, G. I., and Miller, 
K. A., 2003,  MNRAS, 341, 213-229.

\bibitem[]{} Benner, L. and McKinnon, W. 1995.   Icarus, 118,  155.

\bibitem[Benz \& Asphaug(1999)]{1999Icar..142....5B} Benz, W., \& Asphaug, 
E.\ 1999,  Icarus, 142, 5

\bibitem[Boehnhardt et al.(2002)]{2002A&A...395..297B} Boehnhardt, H., et 
al.\ 2002, \aap, 395, 297

\bibitem[Bottke et al.(2005)]{2005Icar..175..111B} Bottke, W.~F., Durda, 
D.~D., Nesvorn{\'y}, D., Jedicke, R., Morbidelli, A., Vokrouhlick{\'y}, D., 
\& Levison, H.\ 2005, Icarus, 175, 111

\bibitem[Burchell \& Johnson(2005)]{2005MNRAS.360..769B} Burchell, M.~J., 
\& Johnson, E.\ 2005,  \mnras, 360, 769

\bibitem[Burns]{} Burns, J. 1986.  In Satellites, edited by J. A. Burns 
and M. S. Matthews, Univ. Az. Press, Tucson, Az.  

\bibitem[Canup \& Ward(2002)]{2002AJ....124.3404C} Canup, R.~M., \& Ward, 
W.~R.\ 2002,  \aj, 124, 3404

\bibitem[Canup \& Ward(2006)]{2006Natur.441..834C} Canup, R.~M., \& Ward, 
W.~R.\ 2006, \nat, 441, 834 

\bibitem[Carpenter et al.(2005)]{2005AJ....129.1049C} Carpenter, J. M., 
Wolf, S., Schreyer, K., Launhardt, R., \& Henning, T.\ 2005,  \aj, 129, 1049 

\bibitem[]{} Carruba, V., Burns, J., Nicholson, P., and Gladman, B. 2002.   
Icarus,
158, 434-449.

\bibitem[]{} Carusi, A. and Valsecchi, G. 1979.   In Asteroids, editor T.
Gehrels, Univ. Az.  Press, Tucson, pp. 391-416.

\bibitem[Chamberlain \& Brown(2004)]{2004Icar..172..163C} Chamberlain, 
M.~A., \& Brown, R.~H.\ 2004, Icarus, 172, 163-169

\bibitem[]{} Christou, A. (2005). Icarus, 174, 215-229.

\bibitem[]{} Clark, R. et al. (2005).  Nature 435, 66-68

\bibitem[]{} Colombo, G., and Franklin, F. A. 1971.   Icarus, 30, 186-189.

\bibitem[]{} Cruikshank, D., Degewij, J., and Zellner, B. 1982.  
In Satellites of Jupiter, ed. D. Morrison,
Univ. Az. Press, Tucson, pp129-146.



\bibitem[]{} Cruikshank, D. 2006.  In Protostars and Planets V, edited by
B. Reipurth, D. Jewitt and K. Keil, Univ. Az. Press, Tucson. (in press)

\bibitem[{\'C}uk \& Burns(2004)]{2004Icar..167..369C} Cuk, M., \& 
Burns, J.~A.\ 2004,  Icarus, 167, 369 

\bibitem[]{} Cuk, M., and Gladman, B. 2006.  Icarus, 183, 362-372.

\bibitem[]{} d'Angelo, G., Henning, T., and Kley, W., 2002, 
Astron. Ap., 385, 647-670.

\bibitem[Dohnanyi(1969)]{1969JGR....74.2431D} Dohnanyi, J.~S.\ 1969, \jgr, 
74, 2431 

\bibitem[Doressoundiram et al.(2002)]{2002AJ....124.2279D} Doressoundiram, 
A., Peixinho, N., de Bergh, C., Fornasier, S., Th{\'e}bault, P., Barucci, 
M.~A., \& Veillet, C.\ 2002, \aj, 124, 2279

\bibitem[Emelyanov(2005)]{2005A&A...438L..33E} Emelyanov, N.~V.\ 2005, 
 \aap, 438, L33-L36.

\bibitem[Fern{\'a}ndez et al.(2003)]{2003AJ....126.1563F} Fern{\'a}ndez, 
Y.~R., Sheppard, S.~S., \& Jewitt, D.~C.\ 2003,  \aj, 126, 1563 - 1574.

\bibitem[Geballe et al.(2002)]{2002Icar..159..542G} Geballe, T.~R., Dalle 
Ore, C.~M., Cruikshank, D.~P., \& Owen, T.~C.\ 2002, 
Icarus, 159, 542 - 544

\bibitem[Gladman et al.(1998)]{1998Natur.392..897G} Gladman, B.~J., 
Nicholson, P.~D., Burns, J.~A., Kavelaars, J.~J., Marsden, B.~G., Williams, 
G.~V., \& Offutt, W.~B.\ 1998, \nat, 392, 897 

\bibitem[]{} Gladman, B., et al. (2000).  Icarus, 147, 320.

\bibitem[]{} Gladman, B. et al.  (2001).   Nature, 412, 163-166.

\bibitem[Goldreich et al.(1989)]{1989Sci...245..500G} Goldreich, P., 
Murray, N., Longaretti, P.~Y., \& Banfield, D.\ 1989,  Science, 245, 500 
 

\bibitem[]{} Grav, T., Holman, M., Gladman, B., and Aksnes, K. (2003).    
Icarus, 166, 33-45.

\bibitem[]{} Grav, T. and Holman, M. (2004).   Ap. J., 605, L141-145.

\bibitem[]{} Grav, T., Holman, M. and Fraser, W. (2004).   Ap. J., 613, L77-80.

\bibitem[]{} Grav, T., and Bauer, J. (2007).   Preprint (astro-ph/0611590)

\bibitem[Hamilton \& Krivov (1997)]{}
Hamilton, D. P., Krivov, A. V., 1997, Icarus, 128, 241-249.

\bibitem[]{} Henon, M. 1970.    Astron. Ap., 9, 24-36.

\bibitem[]{} Heppenheimer, T. A., and Porco, C. C. 1977.  Icarus, 30, 385-401.

\bibitem[]{} Holman, M. et al. (2004).   Nature 430, 
865-867.

\bibitem[Innanen et al.(1997)]{1997AJ....113.1915I} Innanen, K.~A., Zheng, 
J.~Q., Mikkola, S., \& Valtonen, M.~J.\ 1997, \aj, 113, 1915 

\bibitem[]{} Jarvis, K. S. et al. 2000. 
Icarus, 145, 445-453.

\bibitem[Jewitt et al.(2000)]{2000AJ....120.1140J} Jewitt, D.~C., Trujillo, 
C.~A., \& Luu, J.~X.\ 2000, \aj, 120, 1140 

\bibitem[Jewitt(2002)]{2002AJ....123.1039J} Jewitt, D.~C.\ 2002,  
\aj, 123, 1039 - 1049

\bibitem[Jewitt \& Sheppard(2002)]{2002AJ....123.2110J} Jewitt, D.~C., \& 
Sheppard, S.~S.\ 2002,   \aj, 123, 2110

\bibitem[]{} Jewitt, D., Sheppard, S. and Porco, C. 2004.    
In JUPITER, eds. F. Bagenal, T. Dowling and W. McKinnon, Cambridge Univ.
Press, Cambridge. pp. 263-280

\bibitem[]{} Jewitt, D. and Sheppard, S. 2005.  Space Sci. Reviews,  
116, 441-456.

\bibitem[]{} Jewitt, D., Sheppard, S., and Kleyna, J.  2006.    
Scientific American, August issue, pp. 40 - 47.

\bibitem[]{} Johnson, T. and Lunine, J. 2005.  
Nature 435, 69-70.

\bibitem[]{} Kary, D., and Dones, L. 1996.   Icarus 121,
207-224.

\bibitem[]{} Kavelaars, J. et al. 2004.  Icarus, 169, 474-481.

\bibitem[]{} Kessler, D. J. 1981.   Icarus, 48,
39-48.

\bibitem[]{} Kortenkamp, S. 2005.  
Icarus 175, 409-418.

\bibitem[]{} Kozai, Y. 1962.   Astron. J., 67, 591-598.

\bibitem[]{} Krivov, A.~V., Wardinski, I., Spahn, F., Kruger,H. and Grun, E.
2002.  Icarus 157,
436-455.

\bibitem[]{} Kuiper, G. 1956. 
Vistas in Astronomy 2, pp. 1631-1666.  New York, Pergammon.

\bibitem[]{} Kuiper, G. 1961.  In Planets and Satellites,
eds. G. Kuiper and B. Middlehurst, Univ. Chicago Press, Chicago, pp.
575-592.

\bibitem[Lissauer(2005)]{2005SSRv..116...11L} Lissauer, J.~J.\ 2005, Space 
Science Reviews, 116, 11

\bibitem[]{} Lubow, S. H., Seibert, M., and Artymowicz, P., 1999,  ApJ, 
562, 1001-1012.

\bibitem[Lunine \& Stevenson(1982)]{1982Icar...52...14L} Lunine, J.~I., \& 
Stevenson, D.~J.\ 1982, Icarus, 52, 14 - 39.

\bibitem[]{} Luu, J. 1991.    Astron. J., 102, 1213-1225.


\bibitem[Marchis et al.(2006)]{2006Natur.439..565M} Marchis, F., et al.\ 
2006, \nat, 439, 565 

\bibitem[Marzari \& Scholl(1998)]{1998A&A...339..278M} Marzari, F., \& 
Scholl, H.\ 1998, \aap, 339, 278

\bibitem[Mayer et al.(2002)]{2002Sci...298.1756M} Mayer, L., Quinn, T., 
Wadsley, J., \& Stadel, J.\ 2002, Science, 298, 1756

\bibitem[McKinnon \& Leith(1995)]{1995Icar..118..392M} McKinnon, W.~B., \& 
Leith, A.~C.\ 1995,   Icarus, 118, 392

\bibitem[Morbidelli et al.(2005)]{2005Natur.435..462M} Morbidelli, A., 
Levison, H.~F., Tsiganis, K., \& Gomes, R.\ 2005, \nat, 435, 462 

\bibitem[Mosqueira \& Estrada(2003)]{2003Icar..163..198M} Mosqueira, I., \& 
Estrada, P.~R.\ 2003, Icarus, 163, 198 - 231.

\bibitem[]{} Nakamura, T., and Yoshikawa, M. 1995.  
Icarus, vol. 116, p. 113-130.

\bibitem[]{} Nesvorn\'y, D., Alvarellos, J., Dones, L. and Levison, 
H. 2003. Astron. J. 126, 398-429.

\bibitem[]{} Nesvorn\'y, D., Beauge, C., and Dones, L. 2004.   
Astron. J., 127, 1768-1783.

\bibitem[]{} Perrine, C. 1905.  PASP,
17, 22-23.

\bibitem[Pickering(1899)]{1899BHarO..49....1P} Pickering, E. C.\ 1899, 
Harvard College Observatory Bulletin, 49, 1 

\bibitem[]{} Pollack, J. B., Burns, J. A., and Tauber, M. E. 1979.   
Icarus, 37, 587-611.

\bibitem[]{} Pollack, J., Hubickyj, O., Bodenheimer, P., Lissauer, J., 
Podolak, M., and Greenzweig, Y. 1996.   Icarus, 124, 62

\bibitem[Porco et al.(2003)]{2003Sci...299.1541P} Porco, C. C., et al.\ 
2003, Science, 299, 1541 

\bibitem[Porco et al.(2005)]{2005Sci...307.1237P} Porco, C. C., et al.\ 
2005, Science, 307, 1237 

\bibitem[]{} Rettig, T., Walsh, K., and Consolmagno, G.  2001.   
Icarus, 154, 313-320.

\bibitem[Rieke et al.(2005)]{2005ApJ...620.1010R} Rieke, G. H., et al.\ 
2005,  \apj, 620, 1010 

\bibitem[]{} Saha, P., and Tremaine, S. 1993.    Icarus, 106, 549-562.

\bibitem[]{} Sheppard, S. and Jewitt, D. 2003.   Nature, 423, 261-263.

\bibitem[Sheppard et al.(2004)]{2004AJ....128.2542S} Sheppard, S.~S., 
Jewitt, D., \& Kleyna, J. 2004.  \aj, 128, 2542

\bibitem[Sheppard \& Jewitt(2004)]{2004AJ....127.3023S} Sheppard, S. S., 
\& Jewitt, D.\ 2004,  \aj, 127, 3023

\bibitem[]{} Sheppard, S., Jewitt, D. and Kleyna, J. 2005.  
Astron. J., 129, 518-525.

\bibitem[Sheppard et al.(2006)]{2006AJ....132..171S} Sheppard, S. S., 
Jewitt, D., \& Kleyna, J.\ 2006, \aj, 132, 171

\bibitem[Simonelli et al.(1999)]{1999Icar..138..249S} Simonelli, D. P., 
Kay, J., Adinolfi, D., Veverka, J., Thomas, P.~C., \& Helfenstein, P.\ 
1999, Icarus, 138, 249

\bibitem[]{} Smith, D. W., Johnson, P. E., and Shorthill, R. W. 1981.
Icarus, 46, 108-113.

\bibitem[Stansberry et al.(2005)]{2005DPS....37.5205S} Stansberry, J. A., 
Cruikshank, D.~P., Grundy, W.~G., Margot, J.~L., Emery, J.~P., Fern{\'a}ndez, 
Y.~R., \& Rieke, G.~H.\ 2005,  AAS/Division for Planetary Sciences 
Meeting Abstracts, 37.


\bibitem[Stephens \& Noll(2006)]{2006AJ....131.1142S} Stephens, D.~C., \& 
Noll, K.~S.\ 2006, \aj, 131, 1142 
 
\bibitem[]{} Stevenson, D., Harris, A., and Lunine, J. 1986.  Origins of 
Satellites, in  Satellites, eds. J. Burns and M. Matthews, Univ. Az. Press, 
Tucson, pp. 39-88.

\bibitem[Takahashi \& Ip(2004)]{2004PASJ...56.1099T} Takahashi, S., \& Ip, 
W.-H.\ 2004,  \pasj, 56, 1099

\bibitem[]{} Tholen, D. J., and Zellner, B. 1984.   Icarus, 58, 246-253.

\bibitem[Thomas et al.(1991)]{1991JGR....9619253T} Thomas, P., Veverka, J., 
\& Helfenstein, P.\ 1991,  \jgr, 96, 19253 

\bibitem[Touma \& Wisdom(1998)]{1998AJ....115.1653T} Touma, J., \& Wisdom, 
J.\ 1998, \aj, 115, 1653 

\bibitem[Trujillo et al.(2001)]{2001AJ....122..457T} Trujillo, C.~A., 
Jewitt, D.~C., \& Luu, J.~X.\ 2001, \aj, 122, 457

\bibitem[Tsiganis et al.(2005)]{2005Natur.435..459T} Tsiganis, K., Gomes, 
R., Morbidelli, A., \& Levison, H.~F.\ 2005, \nat, 435, 459 

\bibitem[Vieira Neto et al.(2004)]{2004A&A...414..727V} Vieira Neto, E., 
Winter, O.~C., \& Yokoyama, T.\ 2004,  \aap, 414, 727 

\bibitem[Vieira Neto et al.(2006)]{2006A&A...452.1091V} Vieira Neto, E., 
Winter, O.~C., \& Yokoyama, T.\ 2006, \aap, 452, 1091 

\bibitem[Vilas et al.(2006)]{2006Icar..180..453V} Vilas, F., Lederer, 
S.~M., Gill, S.~L., Jarvis, K.~S., \& Thomas-Osip, J.~E.\ 2006, Icarus, 
180, 453 

\bibitem[Weaver et al.(1995)]{1995Sci...267.1282W} Weaver, H.~A., et al.\ 
1995, Science, 267, 1282 

\bibitem[Weidenschilling(2002)]{2002Icar..160..212W} Weidenschilling, 
S.~J.\ 2002, Icarus, 160, 212

\bibitem[]{} Whipple, A., and Shelus, P. 1993.   Icarus, 101. 265-271.

\bibitem[Zahnle et al.(2003)]{2003Icar..163..263Z} Zahnle, K., Schenk, P., 
Levison, H., \& Dones, L.\ 2003,  Icarus, 163, 263

\end{thebibliography}
\end{document}